\documentclass[lettersize,journal]{IEEEtran}
\usepackage{amsmath,amsfonts}
\usepackage{algpseudocode}
\usepackage{tcolorbox}
\usepackage{array}
\usepackage[caption=false,font=normalsize,labelfont=sf,textfont=sf]{subfig}
\usepackage{textcomp}
\usepackage{stfloats}
\usepackage{url}
\usepackage{tabularx}
\usepackage{verbatim}
\usepackage{graphicx}

\usepackage{cite}
\usepackage{bm}

\usepackage[linesnumbered,ruled,vlined]{algorithm2e}
\usepackage{amssymb,stmaryrd}
\usepackage{tikz}
\usepackage{boldline}
\usepackage{makecell}
\usepackage{threeparttable}
\usepackage{multirow}
\usepackage{booktabs}

\usepackage{pgfplots}
\usepgfplotslibrary{groupplots}
\usetikzlibrary{backgrounds}

\usepackage{calc}
\usepackage{colortbl}

\definecolor{TableGray}{gray}{0.9}
\newcommand*\myscale{1}
\newcommand*\circled[2][1]{\renewcommand*\myscale{#1}\tikz[baseline=(char.base)]{
\node[shape=circle,draw,inner sep=1.5pt, scale=\myscale] (char) {#2};}}
\newcommand{\red}[1]{\textcolor{red}{#1}}

\usepackage{balance}
	
\newcommand{\SC}{\text{SC}}
\newcommand{\HRC}{\text{HRC}}
\newcommand{\IRC}{\text{IRC}}
\newcommand{\SSC}{\text{SSC}}
\newcommand{\IME}{\text{IME}}
\newcommand{\UG}{\text{U}}
\newcommand{\tail}{\text{tail}}
\newcommand{\RCA}{\text{RCA}}
\newcommand{\ACE}{\text{ACE}}
\newcommand{\fiveG}{\text{5G}}

\usepackage{amsthm}
\theoremstyle{definition}

\newtheorem{remark}{Remark}
\newtheorem{example}{Example}

\newcommand{\mytextsf}[1]{{\small\textsf{#1}}}
	
\usepackage{siunitx}
	
\begin{document}
\title{Edge-Spreading Raptor-Like LDPC Codes\\for 6G Wireless Systems}
\author{Yuqing Ren,
	Leyu Zhang,
	Yifei Shen,
	Wenqing Song,\\
	Emmanuel Boutillon,
	Alexios Balatsoukas-Stimming,
	and Andreas Burg
	\thanks{Y. Ren, L. Zhang, Y. Shen, W. Song, and A. Burg are with the Telecommunications Circuits Laboratory (TCL), \'{E}cole Polytechnique F\'{e}d\'{e}rale de Lausanne (EPFL), Lausanne 1015, Switzerland (email: \{yuqing.ren, leyu.zhang, yifei.shen, wenqing.song, andreas.burg\}@epfl.ch).}
	\thanks{E. Boutillon is with the Lab-STICC, CNRS UMR 6285, Universit\'{e} Bretagne Sud, 56100 Lorient, France (e-mail: emmanuel.boutillon@univ-ubs.fr).}
	\thanks{A. Balatsoukas-Stimming is with the Department of Electrical Engineering, Eindhoven University of Technology, 5600 MB Eindhoven, The Netherlands (email: a.k.balatsoukas.stimming@tue.nl).}
	\thanks{\emph{Corresponding author: Andreas Burg} (andreas.burg@epfl.ch)}
}


\maketitle

\begin{abstract}
	Next-generation channel coding has stringent demands on throughput, energy consumption, and error rate performance while maintaining key features of 5G New Radio (NR) standard codes such as rate compatibility, which is a significant challenge.
	Due to excellent capacity-achieving performance, spatially-coupled low-density parity-check (SC-LDPC) codes are considered a promising candidate for next-generation channel coding.
	In this paper, we propose an SC-LDPC code family called edge-spreading Raptor-like (ESRL) codes.
	Unlike other SC-LDPC codes that adopt the structure of existing rate-compatible LDPC block codes before coupling, ESRL codes maximize the possible locations of edge placement and focus on constructing an optimal coupled matrix.
	Moreover, a new graph representation called the unified graph is introduced. 
	This graph offers a global perspective on ESRL codes and identifies the optimal edge reallocation to optimize the spreading strategy.
	We conduct comprehensive comparisons of ESRL codes and 5G-NR LDPC codes. 
	Simulation results demonstrate that when all decoding parameters and complexity are the same, ESRL codes have obvious advantages in error rate performance and throughput compared to 5G-NR LDPC codes in some specific scenarios (low and high number of iterations), making them a promising solution towards next-generation channel coding.
\end{abstract}


\section{Introduction}\label{sec:Intro}
\IEEEPARstart{N}{ext-generation} channel coding for enhanced mobile broadband+ (eMBB+) scenarios with high-speed data services targets a peak data rate of~$200$~Gbps to~$1$~Tbps with an energy efficiency target of around~$1$~pJ/bit~\cite{recommendation2023framework}.
This ambitious target represents a~$10\times$ to~$100\times$ more stringent throughput requirements compared to 5G New Radio~(NR)~\cite{5Gstandard2018}.
As standards continue to evolve, channel coding also needs to meet diverse key performance indicators~(KPIs), including higher throughput, enhanced reliability, lower power consumption, reduced computational complexity, rate compatibility, and new capabilities~\cite{zhang2023channel}, leading to a significant~challenge.

As a key technique for eMBB scenarios of 5G-NR~\cite{5Gstandard2016}, low-density parity-check (LDPC) codes~\cite{gallager1962low} have achieved significant success due to capacity-approaching performance~\cite{Urbanke2001TIT} and parallel processing capabilities under belief propagation (BP) decoding~\cite{FR2001FGTIT}.
For finite-length LDPC codes, though BP offers the advantage of reduced decoding complexity, the BP decoding threshold is sub-optimal compared to the maximum~a-posteriori probability (MAP) threshold.
Hence, spatially-coupled LDPC (SC-LDPC) codes (also called LDPC convolutional codes) were proposed in~\cite{Jimenez1999TIT} and~\cite{Kudekar2011TIT}, by superimposing LDPC block codes with a convolutional structure.
In general, block codes have a simple structure for which decoding is easy to implement in hardware, but they have increased complexity and latency when handling frames with large blocklengths. 
In contrast, convolutional codes are effective at decoding long frames but suffer from sequential processing of windowed decoding that introduces decoding latency.
SC-LDPC codes, offering the best of both worlds, achieve the MAP threshold of the underlying block ensembles using BP decoding on a coupled ensemble~\cite{Kudekar2011TIT,kudekar2013spatially}.
Besides, the compact and regular structure of underlying blocks also simplifies the decoding of long frames, and state-of-the-art SC-LDPC decoder implementations have reported high throughput in the range of hundreds of Gbps~\cite{herrmann2021336}.
Note that as stated in~\cite{kurner2022thz}, the information throughput of a decoder depends on clock frequency, achievable parallelism in hardware, code rate, iteration count, and frame length.
In general, clock frequency and hardware parallelism are decided by process technology and area constraints.
Given the slowdown in frequency scaling in advanced technology nodes, extending the frame length becomes a natural choice for~6G to improve peak throughput~\cite{zhang2023channel} and even efficiency.
The adjustable~\emph{coupling length} of SC-LDPC codes is therefore a desirable property for 6G eMBB+ to flexibly adjust and extend the frame length without altering the matrix of underlying codes.
These ideas have attracted widespread interest for 6G eMBB+~\cite{zhang2023channel,schmalen2015spatially}.

For 6G eMBB+, rate compatibility and blocklength scalability are also necessary features for channel coding.
To address the time-varying nature of wireless channels, rate-compatible channel codes must further support incremental redundancy transmission, thereby maximizing the throughput under~\emph{hybrid automatic repeat-request} (HARQ) schemes~\cite{LTE2007HARQ}.
Specifically, a rate-compatible code family is a nested sequence of codes with different rates, where higher rate codes are embedded into lower rate codes~\cite{ha2004rate}.
An initial codeword is transmitted at a high rate, and in case of failed decoding, additional partial redundancy is sent until all redundancy of the mother code is used up.
Numerous works on the construction of rate-compatible LDPC codes have been published such as~\cite{ha2004rate,yazdani2004construction,yue2007design,chen2015protograph,zhang2021protograph,battaglioni2024rate} mainly using~\emph{puncturing} and~\emph{extension}.
The design of such codes can begin with the construction of a relatively small bipartite graph called the~\emph{protograph}~\cite{thorpe2003low}. 
A well-designed protograph applies a~\emph{copy-and-permute} operation (also called~\emph{lifting}) to obtain larger graphs with nodes, thus leading to longer-blocklength LDPC codes~\cite{fossorier2004quasicyclic}. 
In this context, a protograph-based Raptor-like (PBRL) LDPC code family that meets the above requirements is a very popular one~\cite{chen2015protograph}.
Sharing many similarities with LDPC codes, the introduction of Luby transform (LT) codes~\cite{luby2002lt} and Raptor codes~\cite{shokrollahi2006raptor}, which can virtually generate any desired rate, form a novel coding paradigm of \emph{rateless codes}.
Subsequently, PBRL codes restrict the protograph to a Raptor-like structure, achieving both rate compatibility and blocklength adjustment simultaneously. 
Hence, PBRL LDPC codes have been adopted into 5G-NR successfully~\cite{5Gstandard2016}.
The authors of~\cite{li2021construction,zou2024construction} further optimized 5G-NR LDPC codes to enhance the error rate performance for small blocklengths, as they are used for ultra-reliable low-latency communications (URLLC). 
However, throughput and error floor targets of this mode are very different to eMBB+.
In fact, the complexity of BP decoding and hardware increases as the blocklength grows in 5G-NR and for 6G eMBB+.
As indicated by the state-of-the-art 5G-NR LDPC decoder implementation~\cite{ren2024generalized}, the memory overhead required to accommodate the maximum blocklength occupies a significant part of the decoder core area (over~$60\%$) and results in complex routing.
This limits peak throughput to around~$20$~Gbps for 5G-NR LDPC codes, far away from stringent~demands~beyond~5G.
Given the aforementioned advantages of SC-LDPC codes over traditional LDPC block codes, developing a rate-compatible SC-LDPC code family has emerged as a promising solution towards next-generation channel coding for 6G eMBB+.

SC-LDPC codes can be constructed from LDPC block codes using an~\emph{edge-spreading} technique~\cite{mitchell2015spatially}.
Several SC-LDPC code constructions tailored to a specific rate were presented in~\cite{mo2020designing,esfahanizadeh2018finite,mitchell2017edge,battaglioni2017design,battaglioni2019efficient,naseri2021construction,battaglioni2023optimizing,yang2022breaking,roostaie20244,naseri2022average,zhang2016time}, where most of them focus on optimizing edge spreading starting from a given LDPC block code, mitigating the emergence of harmful graph structures.
For instance,~\cite{mo2020designing} proposed a heuristic search combined with a check-and-flip process to construct short-cycle-free SC-LDPC codes by generating multiple auxiliary matrices.
However, the pseudo-random nature of~\cite{mo2020designing} leads to significant complexity, particularly as the matrix size and~\emph{coupling width} increase.
To address this issue, an edge-spreading matrix~\cite{mitchell2017edge} has been employed in recent works~\cite{battaglioni2017design,battaglioni2019efficient,naseri2021construction,battaglioni2023optimizing,yang2022breaking,roostaie20244,naseri2022average} to provide a more compact representation of edge spreading. 
This enables a simple method to identify harmful structures in SC-LDPC codes, regardless of the increase in coupling width.
Naseri~\emph{et al.}~\cite{naseri2021construction} imposed constraints on the edge-spreading matrix to reduce~\emph{dominant trapping sets} and avoid~\emph{error floors},~\cite{battaglioni2023optimizing} developed a strategy to analyze combinations of harmful objects in the matrix,~\cite{yang2022breaking} employed gradient descent to reduce the computational complexity of constructing SC-LDPC codes with a large coupling width, and~\cite{roostaie20244} analyzed the relationship between short cycles and edge spreading to design short-cycle-free SC-LDPC codes with minimal coupling width.
To further incorporate rate compatibility~\cite{si2012rate,zhou2013robust,nitzold2012spatially,wei2013coded,shi2022design}, 
Si~\emph{et al.} introduced the first rate-compatible SC-LDPC code to achieve the capacity of the binary erasure channel (BEC)~\cite{si2012rate}.
Zhou~\emph{et al.}~\cite{zhou2013robust} designed a family of robust rate-compatible punctured SC-LDPC codes by ensuring the recoverability of punctured bits and avoiding harmful graph structures.
Protograph-based rate-compatible SC-LDPC codes were proposed in~\cite{nitzold2012spatially,wei2013coded,shi2022design} following a simple idea.
Namely, a conventional PBRL protograph is first designed, and then the edge-spreading technique is directly applied to construct a coupled code, with both procedures being independent.
Notably,~\cite{shi2022design} achieved performance gains by simply splitting the 5G-NR LDPC matrix as the foundation for SC-LDPC~codes.

However, these SC-LDPC works either focus on optimizing code constructions tailored to a specific rate~\cite{mo2020designing,esfahanizadeh2018finite,mitchell2017edge,battaglioni2017design,battaglioni2019efficient,naseri2021construction,battaglioni2023optimizing,yang2022breaking,roostaie20244,naseri2022average,zhang2016time} or apply the structure of existing rate-compatible LDPC block codes to generate SC-LDPC codes~\cite{nitzold2012spatially,wei2013coded,shi2022design}, thus not exploring the full potential of SC-LDPC codes.
Consequently, there is a need for a holistic design method to construct SC-LDPC codes that simultaneously meet diverse demands~of 6G eMBB+, including rate compatibility, protograph design with coupling, threshold analysis, removal of short cycles, and high throughput potential.
In particular, in high-throughput~6G eMBB+ scenarios, the iteration count significantly impacts throughput, where the constructed SC-LDPC codes need to achieve strong error rate performance within just a few iterations.
It is also necessary to conduct comprehensive comparisons of the designed SC-LDPC codes and standard LDPC codes, especially when all decoding parameters and complexity are consistent.
This would provide a clear assessment of the potential of SC-LDPC codes for next-generation channel coding.
\begin{figure}[t]
	\centering
	\includegraphics[width=\linewidth]{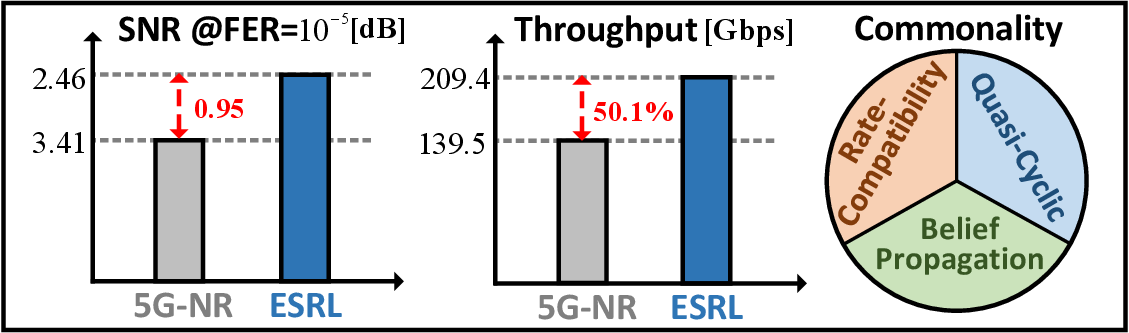}\\
	\caption{Comparisons of ESRL codes and 5G-NR LDPC codes~\cite{5Gstandard2016}. More details are discussed in Section~\ref{sec:SecV_comp}.}\label{fig:CompBar}
\end{figure}

\emph{Contributions:}
This paper proposes an SC-LDPC code family called edge-spreading Raptor-like (ESRL) LDPC codes. 
Our codes target the 6G eMBB+ mode which requires large and scalable transport block sizes with scalable code rates and support for incremental redundancy HARQ (IR-HARQ), making them well-suited for very high-throughput decoding.
Unlike PBRL codes~\cite{chen2015protograph}, ESRL codes maximize the possible locations of edge placement before coupling and focus on constructing an optimal coupled matrix.
Hence, ESRL codes show good performance, while maintaining extensive rate compatibility.
The structure, properties, and code profile of ESRL codes are discussed in~detail.

Moreover, we propose a novel graph representation called the~\emph{unified graph}.
Unlike other edge-spreading optimizations involved with complex processing~\cite{naseri2021construction}, this graph innovatively offers a global perspective on ESRL codes, identifying the optimal edge reallocation to optimize the spreading strategy.
We also introduce modified coupled reciprocal channel approximation (RCA) to analyze the decoding threshold of ESRL codes, demonstrating that ESRL codes enjoy better thresholds across a wide range of rates.
Based on these tools, a complete design flow of ESRL codes is presented.

We also propose a high-throughput decoding algorithm tailored to ESRL codes, called semi-layered multi-engine (SLME) decoding. 
Based on features of ESRL codes, SLME decoding decomposes the original BP decoding into multiple identical sub-decoders that operate in parallel in a more compact manner compared to~\cite{schmalen2015spatially}. 
This approach significantly enhances the decoding parallelism for ESRL~codes.

Finally, we provide a design example of ESRL codes, which targets 6G eMBB+, and conduct comprehensive comparisons with 5G-NR LDPC codes~\cite{5Gstandard2016} and their variants.
As summarized in Fig.~\ref{fig:CompBar}, ESRL codes preserve several advantages and properties of 5G-NR LDPC codes while further enhancing error rate performance and throughput.
When all decoding parameters and complexity are the same, at the rate of~$0.52$, ESRL codes outperform 5G-NR LDPC codes by~$0.95$~dB at five iterations or enhance peak throughput by~$50.1\%$ using comparable hardware complexity~\cite{ren2024generalized}.

\emph{Outline:} The remainder of this paper is organized as follows.
Section~\ref{sec:pre} introduces the symbol definitions and reviews in compact form the preliminaries of LDPC codes.
Section~\ref{sec:ESRL_LDPC} introduces the structure, properties, and code profile of ESRL codes.
Section~\ref{sec:design_and_optimization} outlines the design and optimization procedures for ESRL codes, including the unified graph and modified coupled RCA.
In Section~\ref{sec:punctured_node}, we present the puncturing scheme together with the integration in the IR-HARQ procedure and SLME decoding for ESRL codes.
Section~\ref{sec:design_example} provides a design example of ESRL codes and compares it with 5G-NR LDPC codes.
Finally, Section~\ref{sec:conclusion} concludes~this~paper.
\begin{figure}[t]
	\centering
	\includegraphics[width=\linewidth]{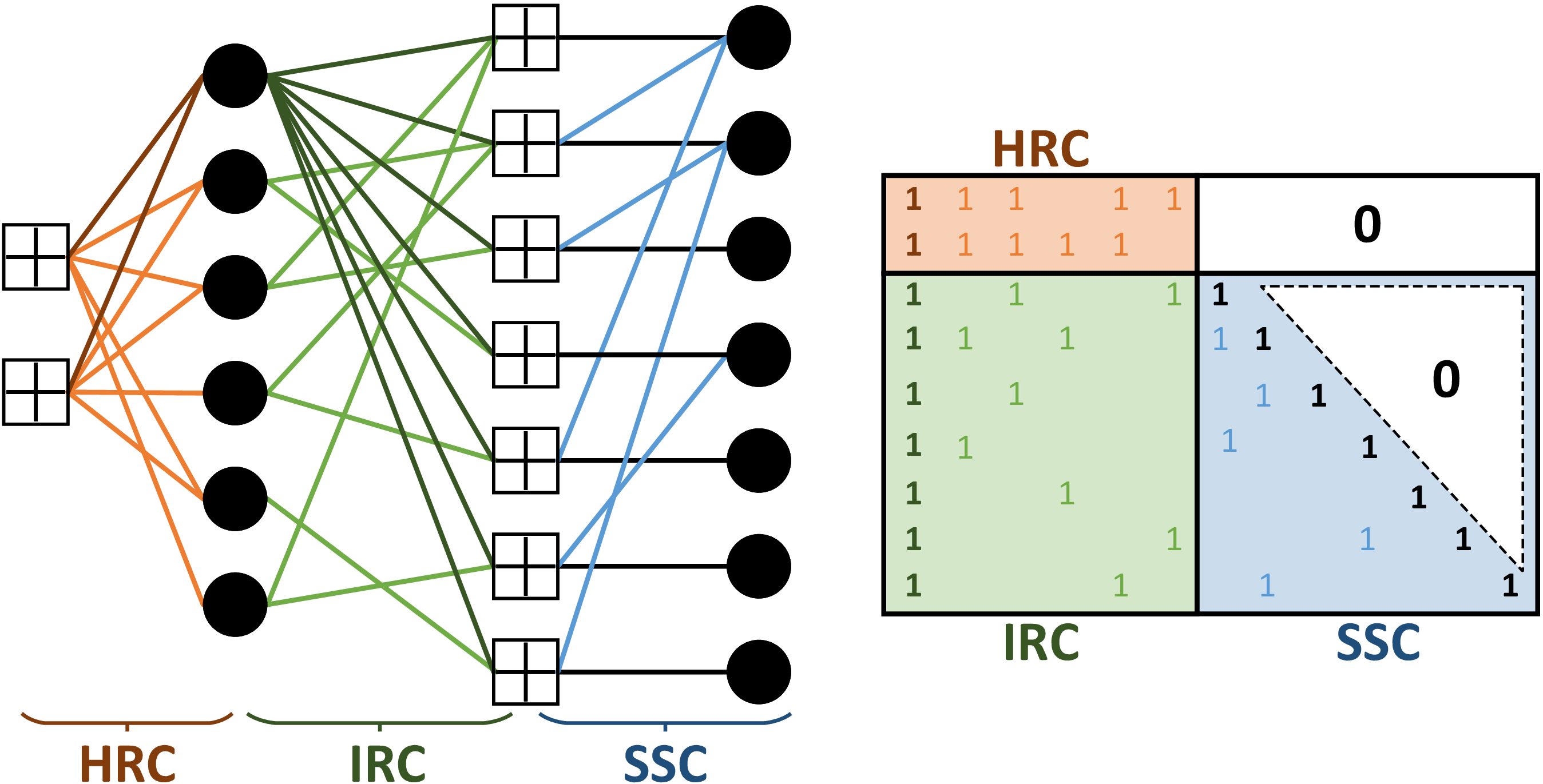}\\
	\caption{Uncoupled protograph and protomatrix for an ESRL code with an HRC, an IRC, and an SSC. The upper right corner is left empty to ensure rate compatibility.}\label{fig:constr}
\end{figure}

\section{Preliminaries}\label{sec:pre}
In this paper, we use the following definitions.
Boldface lowercase letters $\bm{u}$ denote vectors, where $\bm{u}_{i}$ refers to the $i$-th element of $\bm{u}$ and $\bm{u}_{i:j}$ is the sub-vector $[\bm{u}_{i},\bm{u}_{i+1},\dots,\bm{u}_{j}]$,~$i\leq j$, and the null vector otherwise.
Boldface uppercase letters~$\mathbf{B}$ represent matrices, with~$\mathbf{B}_{i,j}$ denoting the element in the $i$-th row of the $j$-th column.
Blackboard letters, such as~$\mathbb{S}\triangleq\{\cdot\}$, denote sets with $|\mathbb{S}|$ being the cardinality.

LDPC block codes are specified by a sparse $M\times N$ parity-check matrix (PCM)~$\mathbf{H}$, which can be further described by a bipartite Tanner graph~\cite{tanner1981recursive} with the information length of~$K=N-M$.
If~$\mathbf{H}_{c,v}=1$ for $0\leq c< M$ and $0\leq v< N$, the $c$-th check node (CN) is connected to the~$v$-th variable node (VN) on the Tanner graph.
We use~$\mathbb{V}_{c}$ to represent the set of adjacent VNs for the $c$-th CN and we use~$\mathbb{C}_{v}$ to denote the set of neighbors for the $v$-th VN.
Let $|\mathbb{V}_c|=d_c$ and $|\mathbb{C}_v|=d_v$ refer to the row degree of the~$c$-th CN and the column degree of the~$v$-th VN, respectively.
Due to the \emph{quasi-cyclic} (QC) property~\cite{fossorier2004quasicyclic}, a QC-LDPC code can be described by a more structured $m\times n$ protomatrix~$\mathbf{B}$ with the information length of~$k=n-m$, where each entry~$\mathbf{B}_{c,v}$ can be extended by a $Z\times Z$ cyclic-shifting identity matrix, with $Z$ being the \emph{lifting size}.
To focus on the protomatrix design in this paper, $\mathbf{B}_{c,v}$ denotes only the number of connected edges between the~$c$-th CN and the~$v$-th VN instead of a specific cyclic shifting value.
For 5G-NR LDPC codes~\cite{5Gstandard2016}, the Raptor-like structure of~$\mathbf{B}$ is shown in~\eqref{eq:NR_B}, which consists of three components to achieve rate compatibility~\cite{chen2015protograph}, i.e., a highest-rate code (HRC)~$\mathbf{B}_{\HRC}$, an incremental redundancy code (IRC)~$\mathbf{B}_{\IRC}$, and an identity~matrix~$\mathbf{I}$.
\begin{equation}\label{eq:NR_B}
	\mathbf{B}=\left[
	\begin{array}{c c}
		\mathbf{B}_{\HRC} & \mathbf{0} \\
		\mathbf{B}_{\IRC} & \mathbf{I} \\
	\end{array}\right].
\end{equation}
Similarly, $\mathbf{B}$ can be transformed into a protograph~$\mathcal{G}$~\cite{thorpe2003low}, defined as $\mathcal{G}\triangleq\{\mathbb{V}\cup\mathbb{C},\mathbb{E}\}$, where $\mathbb{E}$ is the set of all edges in~$\mathcal{G}$.
Let~$e_{c}^{v}$ denote the directed edge from the~$c$-th CN to the~$v$-th VN, and~$e_{v}^{c}$ is the same edge with the opposite direction.
Note that all graph concepts are based on protographs throughout this paper and we omit the term `QC' for simplicity.

Through a convolutional structure, a chain of multiple identical protographs can be organized to derive a protograph-based SC-LDPC code with a coupling width of~$\omega$.
To this end, after edge spreading at each spatial position, $\mathbf{B}$ is split into~$\omega+1$ sub-matrices~$\mathbf{B}_{i}$,~$i\in[0,\omega]$, i.e.,~$\mathbf{B}=\sum_{i=0}^{\omega}\mathbf{B}_i$, each edge in the original~$\mathbf{B}$ is assigned to exactly one of the sub-matrices.
To construct a protomatrix~$\mathbf{B}_{\SC}$ of a finite-length SC-LDPC code, $L$ replicas of a series of sub-matrices~$\mathbf{B}_{i}$,~$i\in[0,\omega]$, are coupled to generate~$\mathbf{B}_{\SC}$ and the corresponding protograph~$\mathcal{G}_{\SC}$, where $L$ is the coupling length.
Consequently, a PCM $\mathbf{H}_{\SC}$ can also be derived from a series of sub-PCMs $\mathbf{H}_{i}$,~$i\in[0,\omega]$ with a lifting size~$Z$.
Let $\bm{c}_{[0,L-1]}\triangleq[\bm{c}_{0},\bm{c}_{1},\dots,\bm{c}_{L-1}]$ denote a codeword of the SC-LDPC code, and $\bm{c}_{[i]}\triangleq[\bm{u}_{[i]},\bm{r}_{[i]}]$ is a partial codeword of the $i$-th block (also called~\emph{batch}) in an IR-HARQ process, where $\bm{u}_{[i]}$ is the information vector and~$\bm{r}_{[i]}$ is the corresponding parity-check vector.
Note that the above $\mathbf{B}_{\SC}$, $\mathcal{G}_{\SC}$, $\mathbf{H}_{\SC}$, and $\bm{c}_{[0,L-1]}$ temporarily ignore the last redundancy tail, which we will explain in detail~below.

\section{Edge-Spreading Raptor-Like LDPC Codes}\label{sec:ESRL_LDPC}
In this section, we introduce the structure, properties, and code profile of the proposed ESRL code family.
\begin{figure*}[t]
	\centering
	\includegraphics[width=\linewidth]{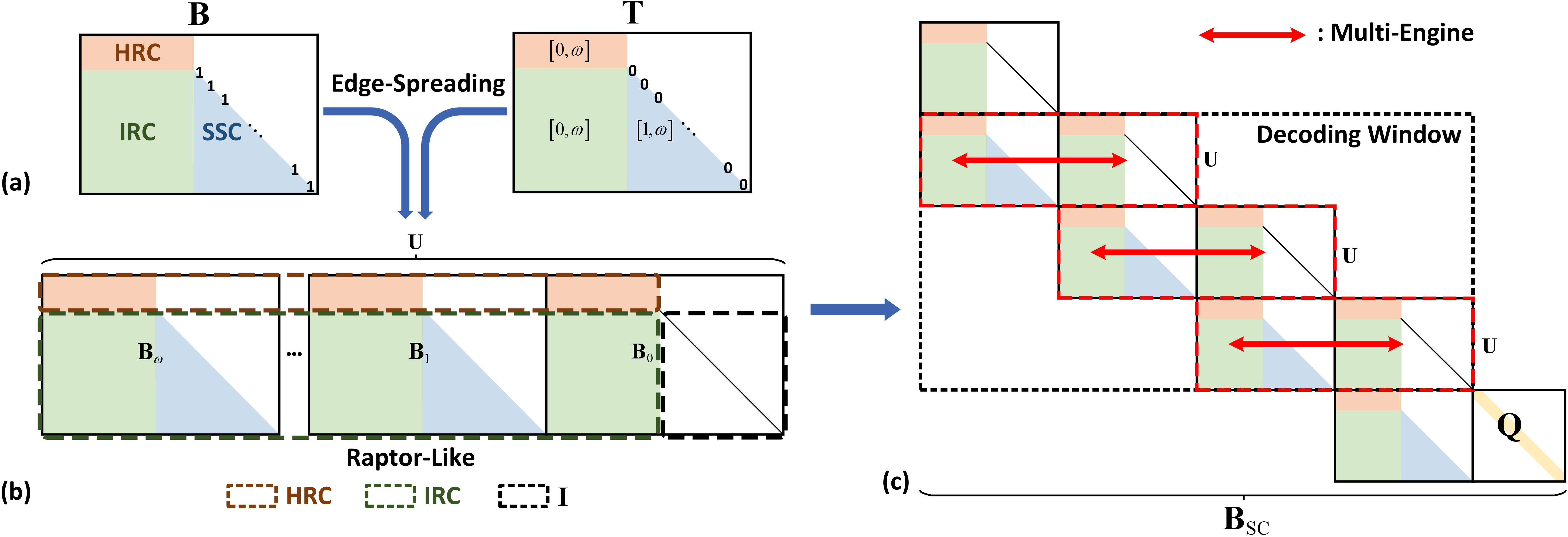}\\
	\caption{The code profile of the ESRL code, including an uncoupled protomatrix~$\mathbf{B}$, an edge-spreading matrix~$\mathbf{T}$, and a tail matrix~$\mathbf{Q}$.}\label{fig:profile}
\end{figure*}

\subsection{Structure of ESRL Codes}\label{sec:SecII-A}
Fig.~\ref{fig:constr} illustrates the structures of the uncoupled protograph~$\mathcal{G}$ and uncoupled protomatrix~$\mathbf{B}$ of an ESRL code, which consists of three components: an HRC~$\mathbf{B}_{\HRC}$, an IRC~$\mathbf{B}_{\IRC}$, and a supplementary-spreading code (SSC)~$\mathbf{B}_{\SSC}$. 
Unlike conventional protographs of PBRL codes~\cite{chen2015protograph}, such as those used in 5G-NR, which include a rightmost identity matrix, an SSC matrix is used to maximize the possible locations for edge placement.
Let~\eqref{eq:ESRL_B} denote~$\mathbf{B}$ of an ESRL code
\begin{equation}\label{eq:ESRL_B}
	\mathbf{B}=\left[
	\begin{array}{c c}
		\mathbf{B}_{\HRC} & \mathbf{0} \\
		\mathbf{B}_{\IRC} & \mathbf{B}_{\SSC} \\
	\end{array}\right].
\end{equation}
Previous approaches to designing rate-compatible SC-LDPC codes~\cite{si2012rate,zhou2013robust,nitzold2012spatially,wei2013coded,shi2022design} typically design a good uncoupled protomatrix first, followed by coupling.
For instance,~\cite{nitzold2012spatially,wei2013coded,shi2022design} attempted to generate rate-compatible SC-LDPC codes by directly applying edge spreading to a PBRL protograph.
However, as shown in Fig.~\ref{fig:profile}, decoding of SC-LDPC codes now occurs on the coupled $\mathbf{B}_{\SC}$ rather than the uncoupled~$\mathbf{B}$ like in LDPC block codes.
This highlights a fundamental difference of ESRL codes compared to~\cite{si2012rate,zhou2013robust,nitzold2012spatially,wei2013coded,shi2022design}, i.e., our target is to design an uncoupled protomatrix~$\mathbf{B}$ that ultimately results in a good coupled protomatrix~$\mathbf{B}_{\SC}$ after edge spreading, regardless of whether~$\mathbf{B}$ by itself is optimal.
Given the Raptor-like structure inherent to ESRL codes, establishing a good Raptor-like structure after edge spreading is clearly more important than good performance before coupling.
With the convolutional structure of SC-LDPC codes, decoding can be further decomposed into multiple identical~\emph{engines} (shown in Fig.~\ref{fig:profile}) working on different~$\mathbf{U}$ blocks at various spatial positions in parallel (also called \emph{multiple-engine decoding} in~\cite{schmalen2015spatially}), where a partially coupled chain~$\mathbf{U}$ is defined as
\begin{equation}\label{eq:ESRL_U}
	\mathbf{U}=\left[\mathbf{B}_{\omega},\dots,\mathbf{B}_{1},\mathbf{B}_{0}\right].
\end{equation} 
Intuitively, decoding on~$\mathbf{U}$ largely determines the overall performance of SC-LDPC decoding. 
However, a conventional PBRL structure restricts the design space to find the optimal~$\mathbf{B}_{\SC}$, since its rightmost SSC is only an identity matrix that must be fully allocated to~$\mathbf{B}_{0}$ for encoding~\cite{pusane2008implementation}. 
This reduces the possible locations for edge placement and leaves the lower triangular parts of $\mathbf{B}_{\omega},\mathbf{B}_{\omega-1},\dots,\mathbf{B}_{1}$ in~\eqref{eq:ESRL_U}~empty. 
The proportion of empty parts increases to $67.6\%$ across medium to low rates, leading to performance loss~\cite{nitzold2012spatially,wei2013coded,shi2022design}.
Hence, SC-LDPC codes need a new structure for the uncoupled protomatrix to enhance message propagation on~$\mathbf{U}$.
Given that~\cite{ha2004rate,yazdani2004construction,yue2007design} have indicated that rate-compatible LDPC codes must adhere to a lower triangular constraint on the right side, either uncoupled or coupled, the ESRL code also follows this constraint to preserve the upper right corner of~\eqref{eq:ESRL_B} empty and keep the SSC lower triangular.
Note that despite the motivation of ESRL codes to design the optimal~$\mathbf{B}_{\SC}$ after edge spreading, the ESRL code still retains the fundamental HRC and IRC parts of traditional PBRL codes.
This means an ESRL decoder can be naturally compatible with PBRL codes of appropriate sizes such as 5G-NR LDPC codes.

The ESRL structure also resembles Raptor codes~\cite{shokrollahi2006raptor}, but with key differences.
First, while the HRC is identical to the precode in a Raptor code, the ESRL code is systematic, with the codeword of each batch represented as $\bm{c}_{[i]}=[\bm{u}_{[i]},\bm{r}_{[i]}],0\leq i<L$.
This design allows explicit transmission of the HRC VNs in all batches and supports decoding at the highest rate, which is infeasible in a Raptor code.
Second, similar to the LT component in Raptor codes, XOR operations in the ESRL code generate the VNs of the IRC and the SSC, producing additional parity bits row-wise.
However, these connections in the ESRL code are pre-defined rather than arbitrary.
Third, compared to two-phase BP decoding of Raptor codes or joint BP decoding on PBRL codes, the ESRL code can support a more flexible windowed BP decoding to slide over the coupled protograph $\mathcal{G}_{\SC}$ in a low-complexity manner~\cite{iyengar2011windowed}.
More details on the decoding are discussed in Section~\ref{subsec:decoding} and Section~\ref{sec:winBPfullBP}.

\subsection{Properties of ESRL Codes}\label{sec:SecII-B}
In this section, we present some important properties of ESRL codes. 
First, the ESRL code is a time-invariant SC-LDPC code with no parallel edges, i.e., each entry of~$\mathbf{B}$ is binary, and no non-zero entries in~$\mathbf{B}_{i},\forall i\in[0,\omega]$ occupy the same position.
The time-invariant nature and the absence of parallel edges ensure that the ESRL code can be decomposed into multiple independent identical engines that decode on different~$\mathbf{U}$ blocks in parallel, without causing memory access conflicts (assuming the decoding schedule is consistent at each engine). 
This results in ESRL codes that allow for significant decoding parallelism to potentially achieve the high throughput demand of next-generation channel coding, as demonstrated in Section~\ref{subsec:decoding} and Section~\ref{sec:design_example}.
It is noteworthy that although parallel edges can reduce the decoding threshold~\cite{chen2015protograph,divsalar2009capacity} and can be easily assigned to $\omega$ sub-matrices in SC-LDPC codes without the need for two-step lifting~\cite{divsalar2009capacity}, this performance gain sacrifices parallelism in an ESRL decoder.
The absence of parallel edges, as mentioned in~\cite{naseri2022average}, can also simplify the theoretical tools described in Section~\ref{sec:design_and_optimization} to optimize~ESRL~codes.

The HRC protograph is the protograph design of a good code at a given high rate, which has been well studied in~\cite{divsalar2009capacity}. 
Following the example of~\cite{divsalar2009capacity}, we set the minimum column weight to at least $3$ to ensure linear minimum distance growth with increasing blocklength.
Besides, a punctured VN that is fully connected to all CNs in the HRC is placed in the first column to reduce the decoding threshold.
The IRC component aims to provide reliability back to the VNs of the original HRC. 
In a complete ESRL matrix with the lowest rate, the CNs of the IRC eventually connect back to all VNs of the HRC.
Following~\cite{garcia2003approaching}, each row of the IRC has at least two edges to retain linear minimum distance growth with increasing blocklength.
Besides, the punctured VN of the precode connects to all CNs in the IRC.
Unlike the use of parallel edges in~\cite{chen2015protograph}, the HRC and the IRC of ESRL codes do not involve any parallel edges.

The SSC protograph is to enhance message propagation on~$\mathbf{U}$, especially for medium to low rates.
In the SSC, all diagonal elements are set to `$1$'s to facilitate encoding.
Note that the SSC is connected to parity bits that are generally less critical than the VNs in the HRC, thus the number of edges in the SSC should be kept lower than that of the IRC.
While the rate reduces, we increase the SSC weight to avoid the right side being too sparse, but by experimental trial, this proportion is at most $30\%$ of the IRC weight.

\subsection{Code Profile of ESRL Codes}\label{sec:SecII_codeprofile}
In this section, we present the code profile of the ESRL code family. 
Compared with LDPC block codes, SC-LDPC codes contain more information, such as a coupling relationship, which cannot be described by a single uncoupled protomatrix~$\mathbf{B}$.
Although a straightforward representation using the coupled protomatrix~$\mathbf{B}_{\SC}$ is feasible, such a description would contain a significant amount of redundant information due to the presence of~$L$ replicas of~$\mathbf{U}$.
In this paper, we provide a simpler method to represent ESRL codes. 
Namely, a complete ESRL code can be described using three matrices: an uncoupled protomatrix~$\mathbf{B}$, an edge-spreading matrix~$\mathbf{T}$, and a tail matrix~$\mathbf{Q}$, as illustrated~in~Fig.~\ref{fig:profile}(c).

The uncoupled protomatrix~$\mathbf{B}$ outlines the parent structure of ESRL codes, as explained in Section~\ref{sec:SecII-A}.
To convert~$\mathbf{B}$ to~$\mathbf{B}_{\SC}$, we adopt an edge-spreading matrix~$\mathbf{T}$ that was proposed in~\cite{mitchell2017edge} and has been adopted in~\cite{naseri2021construction,naseri2022average}, instead of explicitly recording $\omega+1$ sub-matrices as in~\cite{mitchell2015spatially,mo2020designing}.
The matrix~$\mathbf{T}$ of size~$m\times n$ determines how the edges of the uncoupled protograph~$\mathcal{G}$ are spread to the coupled protograph~$\mathcal{G}_{\SC}$.
Specifically, the elements of~$\mathbf{T}$ indicate to which sub-matrices the edges of~$\mathbf{B}$ are assigned.
We set $\mathbf{T}_{c,v}=-1$ if $\mathbf{B}_{c,v}=0$.
Otherwise, $\mathbf{T}_{c,v}=i$, $i\in[0,\omega]$, implies that the entry at row $c$ and column $v$ of~$\mathbf{B}_{c,v}$ is assigned to the corresponding position in~$\mathbf{B}_{i}$.
Let~$\delta_i(\mathbf{T}_{c,v})$ be a function that returns a boolean value of `$1$' if~$\mathbf{T}_{c,v}$ equals $i$, and `$0$' otherwise.
As a result, we can generate $\omega+1$ sub-matrices~$\mathbf{B}_{0},\mathbf{B}_{1},\dots,\mathbf{B}_{\omega}$ explicitly as
\begin{equation}\label{eq:Bi}
	\mathbf{B}_i\!\!=\!\!\left[\!\!\!\begin{array}{c c c}
		\mathbf{B}_{0,0}\cdot\delta_i(\mathbf{T}_{0,0}) & \cdots & \mathbf{B}_{0,n-1}\cdot\delta_i(\mathbf{T}_{0,n-1})\\
		\vdots & \ddots & \vdots\\
		\mathbf{B}_{m\!-\!1,0}\!\cdot\!\delta_i(\mathbf{T}_{m\!-\!1,0})\!\!\!\!\!& \cdots & \!\!\!\!\mathbf{B}_{m\!-\!1,n\!-\!1}\!\cdot\!\delta_i(\mathbf{T}_{m\!-\!1,n\!-\!1})\\
	\end{array}\!\!\!\right].
\end{equation}

Furthermore, the termination of finite-length SC-LDPC codes results in an unavoidable rate loss at the boundary.
To mitigate this issue, one can either append a redundancy matrix~$\mathbf{Q}$ at the end~\cite{SC_LDPC_database,pusane2008implementation} or reduce the information length of the last block~\cite{graell2020forward} to satisfy encoding. 
ESRL codes adopt the former approach, as it has a lower rate loss.
Previous works generally use a simple identity matrix for~$\mathbf{Q}$~\cite{SC_LDPC_database,pusane2008implementation}, but this leads to serious error floors at medium to low rates.
In contrast, we adopt a tri-diagonal matrix for~$\mathbf{Q}$ with a lower triangular structure to ensure that most column degrees are either~$2$ or~$3$.
For rate compatibility, the matrix~$\mathbf{Q}$ must remain full-rank even after pruning.
Based on the three matrices~$\mathbf{B}$, $\mathbf{T}$, and~$\mathbf{Q}$, we explicitly define the coupled protomatrix~$\mathbf{B}_{\SC}$ of the ESRL code as follows
\begin{equation}\label{eq:Bsc}
	\begingroup
	\renewcommand{\arraystretch}{0.5}
	\mathbf{B}_{\SC}\!=\!\left[\!\!
	\begin{array}{ccccccc}
		\mathbf{B}_{0}      &                     &        &                     &                & \cdots & \mathbf{0}\\
		\mathbf{B}_{1}      & \mathbf{B}_{0}      &        &                     &                &        & \vdots\\
		\vdots              & \mathbf{B}_{1}      & \ddots &                     &                &        & \\
		\mathbf{B}_{\omega} & \vdots              & \ddots & \mathbf{B}_{0}      &                &        & \\
		                    & \mathbf{B}_{\omega} & \ddots & \mathbf{B}_{1}      & \mathbf{Q}_{1} & \cdots & \mathbf{0}\\
		\vdots              &                     & \ddots & \vdots              & \vdots         & \ddots & \vdots\\
		\mathbf{0}          & \cdots              &        & \mathbf{B}_{\omega} & \mathbf{0}     & \cdots & \mathbf{Q}_{\omega}\\
	\end{array}\!\!\right]\!.
	\endgroup
\end{equation}
Without loss of generality, we decompose the tail matrix~$\mathbf{Q}$ in~\eqref{eq:Bsc} into a series of matrices $\mathbf{Q}_{1},\dots,\mathbf{Q}_{\omega}$ (each of size~$m\times m$) on the diagonal, where an example of~$\mathbf{Q}$ is illustrated in Fig.~\ref{fig:exampleDesign}.
It is noteworthy that the three matrices~$\{\mathbf{B},\mathbf{T},\mathbf{Q}\}$ allow for flexible pruning from the end, enabling the ESRL code to support a wide range of rates with one code matrix.
Let $\rho$ denote the number of punctured VNs in $\mathbf{B}$, we can define the code rate~$R$ of the ESRL code as
\begin{equation}\label{eq:Rsc}
	R=\frac{k\cdot L}{n\cdot L+m\cdot\omega-\rho\cdot(L+\omega)}.
\end{equation}

\section{Design of ESRL Codes}\label{sec:design_and_optimization}
In this section, we outline the procedures for the design and optimization of a good ESRL code in detail.

\subsection{Unified Graph for ESRL Codes}\label{subsec:unifiedGraph}
\begin{figure}[t]
	\centering
	\includegraphics[width=\linewidth]{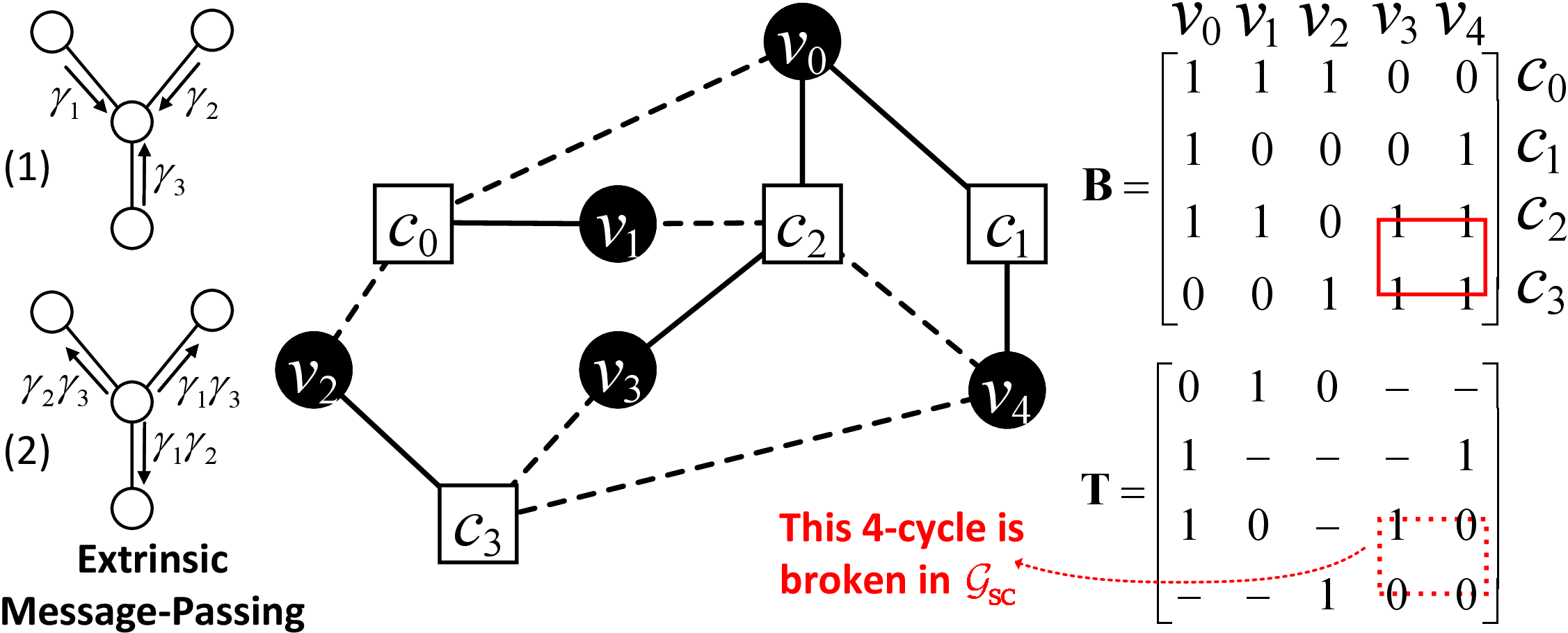}\\
	\caption{A unified graph~$\mathcal{G}_{\UG}$ example, where dashed black lines and solid black lines are assigned to $\mathbf{B}_0$ and $\mathbf{B}_1$, respectively.}\label{fig:uniGraph}
\end{figure}

Inspired by the aforementioned code profile, we propose a novel graph representation for ESRL codes, which we call the unified graph~$\mathcal{G}_{\UG}$. 
This graph contains all essential code information and enables a comprehensive illustration of possible graph structures within an SC-LDPC protograph. 
Unlike traditional approaches, a single unified graph~$\mathcal{G}_{\UG}$ can integrate the relationships and dependencies among the matrices~$\mathbf{B}$ and~$\mathbf{T}$ in ESRL codes,\footnote{Given that the tail matrix~$\mathbf{Q}$ has a fixed structure and is not involved in edge spreading, we ignore the presence of~$\mathbf{Q}$ to simplify the optimization on~$\mathcal{G}_{\UG}$. Note that $\mathcal{G}_{\UG}$ is a protograph concept.} thereby offering a more global view of the code construction.
The construction of~$\mathcal{G}_{\UG}$ is built on the bipartite graph derived from~$\mathbf{B}$, where edges are annotated by different line types based on the corresponding elements of~$\mathbf{T}$.
For example, Fig.~\ref{fig:uniGraph} depicts the unified graph~$\mathcal{G}_{\UG}$ derived from~$\mathbf{B}$ and~$\mathbf{T}$ with $\omega=1$, where dashed and solid lines distinguish whether edge $e$ is assigned to $\mathbf{B}_0$ or $\mathbf{B}_1$, respectively.
In this section, we present a message-passing algorithm on~$\mathcal{G}_{\UG}$ to count and eliminate short cycles of~$\mathcal{G}_{\SC}$ efficiently, which inspires a good spreading strategy on~$\mathcal{G}_{\UG}$ to enhance the performance~of~ESRL~codes.

\subsubsection{Cycle Counting over the Unified Graph}
We first define cycles of length~$l$ (referred to as~\emph{$l$-cycles}) in a protograph~$\mathcal{G}$.
An $l$-cycle corresponding to a sequence of~$l/2$ VNs and~$l/2$ CNs~in~\eqref{eq:unigraph_cycle}, or equivalently, a sequence of~$l$~edges~in~\eqref{eq:unigraph_cycle_e}, starts and ends at the same node.
\begin{equation}\label{eq:unigraph_cycle}
	\{v_0,c_0,v_1,c_1,\ldots,v_{l/2-1},c_{l/2-1},v_0\}.
\end{equation}
\begin{equation}\label{eq:unigraph_cycle_e}
	\{e_{v_{0}}^{c_{0}},e_{c_{0}}^{v_{1}},e_{v_{1}}^{c_{1}},e_{c_{1}}^{v_{2}},\dots,e_{v_{l/2-1}}^{c_{l/2-1}},e_{c_{l/2-1}}^{v_{0}}\}.
\end{equation}
For rigor, we only consider \emph{simple cycles} in this paper, i.e., \emph{tailless backtrackless closed (TBC) walks} in a graph.
After edge spreading from~$\mathcal{G}$ to~$\mathcal{G}_{\SC}$, the number of $l$-cycles in~$\mathcal{G}$ is the upper bound of the number of~$l$-cycles in~$\mathcal{G}_{\SC}$ if repeated cycles in~$L$ replicas are counted only once~\cite{naseri2022average}.
\begin{proof}
	Let~$\phi$ denote the accumulated path value for an~$l$-cycle, as calculated by
	\begin{equation}\label{eq:unigraph_TBC_T}
		\phi=\Delta \mathbf{T}_0+\Delta \mathbf{T}_1+\Delta \mathbf{T}_2+\ldots+\Delta \mathbf{T}_{l/2-1},
	\end{equation}
	where~$\Delta\mathbf{T}_{i}=\mathbf{T}_{c_i,v_i}-\mathbf{T}_{c_i,v_{\text{mod}(i+1,l/2)}}$.
	Note that~$\phi$ tracks the path of how an $l$-cycle passes through~$\omega+1$ sub-matrices in SC-LDPC codes.
	Since an~$l$-cycle in~$\mathcal{G}$ remains in~$\mathcal{G}_{\SC}$ only if $\phi=0$~\cite{mitchell2017edge,naseri2022average} otherwise broken,~$\mathcal{G}_{\SC}$ will not generate any new unique cycle compared~to~$\mathcal{G}$.
\end{proof}

Subsequently, we introduce a message-passing algorithm over~$\mathcal{G}_{\UG}$ to efficiently count cycles for ESRL codes.
In general, a message-passing algorithm computes messages at the nodes and propagates them along the edges of a graph, where the calculation of the message sent along an edge $e$ excludes the message previously received along $e$ (refers to this property as \emph{extrinsic})~\cite{FR2001FGTIT}.
Let~$\gamma$ denote a basic message, and an example of an extrinsic message-passing process is illustrated in Fig.~\ref{fig:uniGraph}.
Hence, the messages over the edges of~$\mathcal{G}_{\UG}$ are all monomials of~$\gamma$ and the message-passing operation is multiplication.
Besides, we need to record the extra accumulated path value~$\phi$ of each~$\gamma$ over~$\mathcal{G}_{\UG}$, due to using edge spreading.
Let~$t$ denote the index of discrete time, and we call each~$t$ as one step, corresponding to either a VN-to-CN or CN-to-VN phase, where~$t$ always starts from~$0$ of the VN-to-CN phase.
We denote~$\mathcal{M}_{v,c}^{(t)}$ as the VN-to-CN message along~$e_{v}^{c}$ and~$\mathcal{M}_{c,v}^{(t)}$ as the CN-to-VN message along~$e_{c}^{v}$.
The update rules are defined in~\eqref{eq:unigraph_v2c} and~\eqref{eq:unigraph_c2v}, and the~$\phi$ of each message is also tracked dynamically to verify whether the cycle remains~or~not~in~$\mathcal{G}_{\SC}$ based on~\eqref{eq:unigraph_TBC_T}.

Herein, we present how to use~$\mathcal{G}_{\UG}$ count cycles that include a specific VN.
For the~$v$-th VN, the initial message~$\mathcal{M}_{v,c}^{(0)}$ at~$t=0$ along the edge~$e_{v}^{c}$ is denoted by~$\gamma_{c}^{\phi_c}$, where~$\phi_c$ is initialized as~$\mathbf{T}_{c,v}$.
We can summarize the VN-to-CN message-passing rules~as
\begin{equation}\label{eq:unigraph_v2c}
	\mathcal{M}_{v,c}^{(t)} = \prod\limits_{c'\in\mathbb{C}_v\backslash c} h^+\left(\mathcal{M}_{c',v}^{(t-1)},\mathbf{T}_{c',v}\right),\quad t\geq 1.
\end{equation}
Similarly, the CN-to-VN message-passing follows
\begin{equation}\label{eq:unigraph_c2v}
	\mathcal{M}_{c,v}^{(t)} = \prod\limits_{v'\in\mathbb{V}_c\backslash v} h^-\left(\mathcal{M}_{v',c}^{(t-1)},\mathbf{T}_{c,v'}\right),\quad t\geq 1.
\end{equation}
The $h^{+}(\cdot)$ and $h^{-}(\cdot)$ functions are defined as~\eqref{eq:unigraph_g1} and~\eqref{eq:unigraph_g2} to update the accumulated path value~$\phi$ in the current message.
Let~$\Phi(\cdot)$ denote a function to extract the~$\phi$ from all~$\gamma$ variables.
\begin{equation}\label{eq:unigraph_g1}
	h^+(\mathcal{M}_{c,v},\mathbf{T}_{c,v})   = \Phi(\mathcal{M}_{c,v})+\mathbf{T}_{c,v}.
\end{equation}
\begin{equation}\label{eq:unigraph_g2}
	h^-(\mathcal{M}_{v,c},\mathbf{T}_{c,v})   = \Phi(\mathcal{M}_{v,c})-\mathbf{T}_{c,v}.
\end{equation}
Notably, the general message format~$\mathcal{M}$ that propagates along the edges in~$\mathcal{G}_{\UG}$ is the product of multiple $\gamma$ variables, each labeled by individual path value~$\phi$, where both the number and type of~$\gamma$ are dynamic.
At the~$t\!=\!l-1$-th step, we will obtain the number of $l$-cycles that include the $v$-th VN~as
\begin{equation}\label{eq:unigraph_NumCycle}
	Q_{v}^{l} = f\left(p_{v}^{l})\right/2,
\end{equation}
where the division by~$2$ arises from the directed edges, as each cycle is counted twice.
The function~$f\left(p_{v}^{l}\right)$ computes the number of all~$\gamma$ variables with~$\phi=0$ in the polynomial~$p_{v}^{l}=\sum_{c\in\mathbb{C}_{v}} \mathcal{M}_{c,v}^{(l-1)}$, i.e., only sequences with $\phi=0$ will remain as cycles in~$\mathcal{G}_{\SC}$ after edge spreading.
Note that due to the extrinsic property mentioned before, the component formed by the initial message~$\gamma_{c}$ sent along the edge~$e_{v}^{c}$ should be excluded from $\mathcal{M}_{c,v}^{(l-1)}$ in~\eqref{eq:unigraph_NumCycle}~\cite{karimi2012message}.
Consequently, we repeat the above procedures for all VNs.
The total number of $l$-cycles in the entire~$\mathcal{G}_{\SC}$ can be obtained as
\begin{equation}\label{eq:unigraph_NumCycleAll}
	Q^{l} = \left(\sum\limits_{v\in\mathbb{V}} Q_v^{l}\right)/\left(l/2\right),
\end{equation} 
where the accumulated result in~\eqref{eq:unigraph_NumCycleAll} needs to be divided by $l/2$ as every $l$-cycle is counted $l/2$-times.
Define the shortest cycle length in a graph as the~\emph{girth}~$g$.
The above approach can efficiently count~$l$-cycles when~$l\in\{4,6,\dots,2g-2\}$.
\begin{figure}[t]
	\centering
	\includegraphics[width=\linewidth]{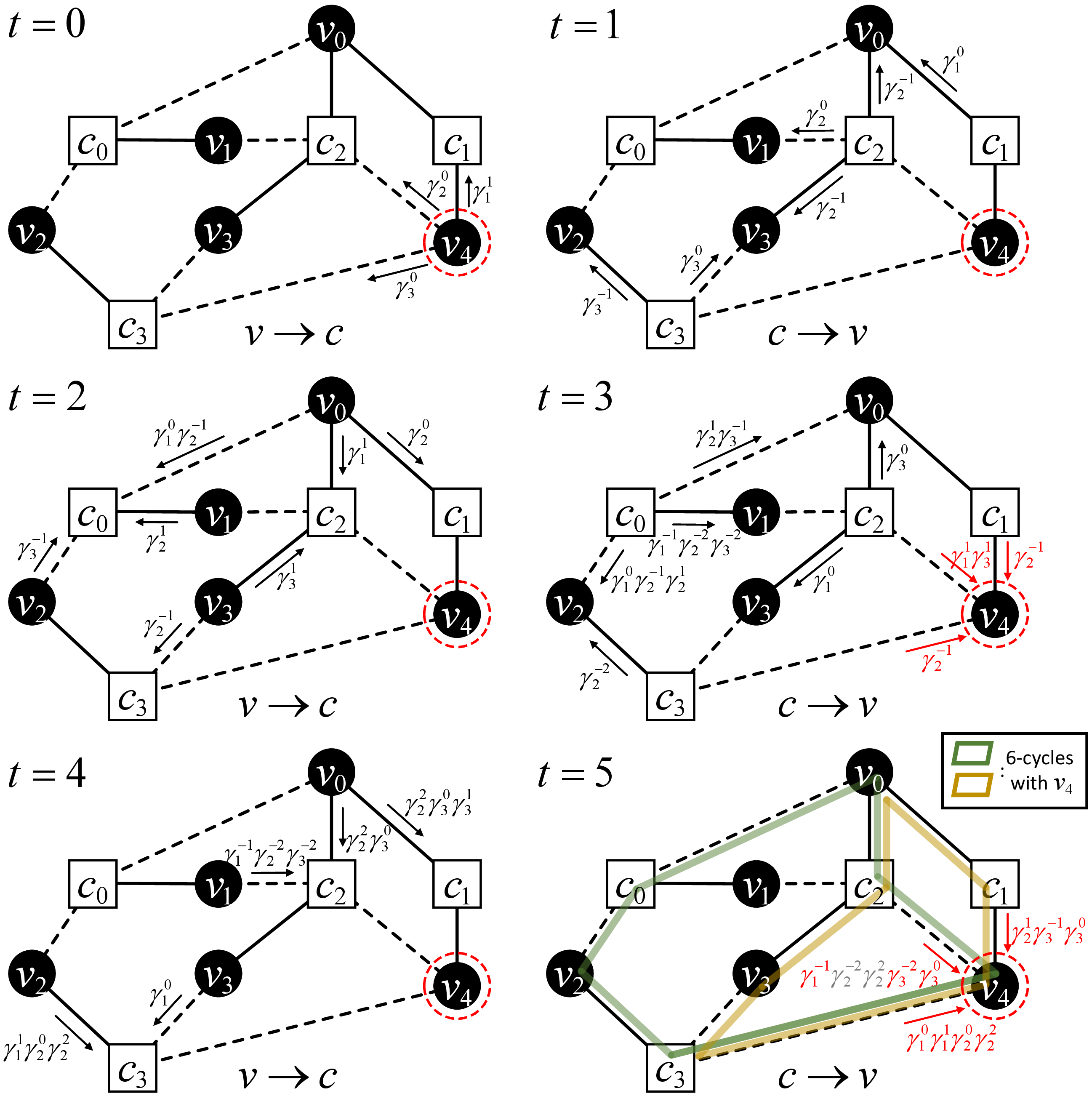}\\
	\caption{An example of cycle counting on a unified graph~$\mathcal{G}_{\UG}$.}\label{fig:uniGraph_example}
\end{figure}

\begin{example}
	We provide an example in Fig.~\ref{fig:uniGraph_example}, illustrating the use of the message-passing algorithm over~the unified graph~$\mathcal{G}_{\UG}$ to count~$4$-cycles and~$6$-cycles~containing~$v_4$.
	\begin{enumerate}
		\setlength{\itemsep}{0pt}
		\setlength{\parsep}{0pt}
		\setlength{\parskip}{0pt}
		\item At~$t=0$, $v_4$ sends the initial messages~$\gamma_{1}^{1}$,~$\gamma_{2}^{0}$, and~$\gamma_{3}^{0}$ along the adjacent edges~$e_{v_4}^{c_1}$,~$e_{v_4}^{c_2}$, and~$e_{v_4}^{c_3}$, respectively, where the initial path values are according to the corresponding elements of~$\mathbf{T}$ as~$\{1,0,0\}$.
		All the messages sent by~$v_{0}$, $v_{1}$, $v_{2}$, and~$v_{3}$ along their edges are equal to $1$ and omitted.
		At~$t=1$, all valid messages are updated based on~\eqref{eq:unigraph_c2v}~and~\eqref{eq:unigraph_g2} for the CN-to-VN~phase.
		
		\item At~$t=2$, all VNs have valid incoming messages, which then pass extrinsic messages using multiplication of~\eqref{eq:unigraph_v2c} and update~$\phi$ by~\eqref{eq:unigraph_g1}.
		At~$t=3$, based on~\eqref{eq:unigraph_NumCycle} we obtain the number of~$4$-cycles as $Q_{v_4}^{4}=f\left(\gamma_2^{-1}+\gamma_1^{1}\gamma_3^{1}+\gamma_2^{-1}\right)/2=(0+0+0)/2=0$.
		
		\item At $t=4$, all VNs propagate messages to their neighbors.
		For simplicity, we omit the messages that fail to return to~$v_4$ at $t=5$ in the graph.
		Note that~$t=5$, $\gamma_{2}^{-2}\gamma_{2}^{2}$ needs to be removed from $\mathcal{M}_{c_2,v_4}^{(5)}$.
		Since the~$\gamma_2$ variable is related to the initial message that passes along an edge and finally returns to~$v_4$ by the same edge, it will not indicate a valid cycle.
		Therefore, we have $Q_{v_4}^{6}=f\left(p_{v_4}^{6}\right)=f\left(\gamma_2^1\gamma_3^{-1}\gamma_3^0+\gamma_1^{-1}\gamma_3^{-2}\gamma_3^{0}+\gamma_1^{0}\gamma_1^1\gamma_2^0\gamma_2^2\right)/2=(1+1+2)/2=2$.
	\end{enumerate}
\end{example}

Hence, this example illustrates that there are no $4$-cycles but two $6$-cycles passing through~$v_4$ in~$\mathcal{G}_{\UG}$.

\subsubsection{Spreading Optimization over the Unified Graph}
In this section, we present how to design a good edge-spreading strategy on the unified graph~$\mathcal{G}_{\UG}$.
When we have utilized~$\mathcal{G}_{\UG}$ to explicitly count the cycles associated with each VN and the total number of cycles in~$\mathcal{G}_{\SC}$, $\mathcal{G}_{\UG}$ can further guide us in optimizing the spreading matrix to~minimize~short~cycles.

Unlike~\cite{battaglioni2019efficient,esfahanizadeh2018finite}, there is no need to repeatedly run the above message-passing algorithm on~$\mathcal{G}_{\UG}$ to update the cycle distribution when changes occur in~$\mathcal{G}_{\UG}$. 
Instead, we can simply adjust the polynomial~$p_v$ in~\eqref{eq:unigraph_NumCycle} to measure the impact of reallocating a specific edge on the cycle distribution.
It is noteworthy that the extra complexity of such an optimization on~$\mathcal{G}_{\UG}$ is negligible for a small~$\omega$, such as the frequent choice of~$\omega=1$.
Specifically, if the edge~$e_{v}^{c}$ is re-assigned from~$\mathbf{T}_{c,v}$ to~${\widetilde{\mathbf{T}}_{c,v}}$ (where ${\widetilde{\mathbf{T}}_{c,v}}\in[0,\omega]$ and ${\widetilde{\mathbf{T}}_{c,v}}\neq\mathbf{T}_{c,v}$), let $\theta={\widetilde{\mathbf{T}}_{c,v}}-\mathbf{T}_{c,v}$, we modify the original~$p_v$ by two steps: \circled[0.8]{1}~Add~$\theta$ to all accumulated path values~$\phi$ of the initial~$\gamma_c$ variable in the polynomial~$p_{v}$ \circled[0.8]{2}~Subtract~$\theta$ from the~$\phi$ of all~$\gamma$ variables in the monomial passed to the VN~$v$ along the edge~$e_c^v$ (as both $e_{v}^c$ and $e_{c}^{v}$ belong to the same undirected edge, they are equivalent in the spreading process).

Following the example of Fig.~\ref{fig:uniGraph_example}, we observe that~$v_4$ is involved in no $4$-cycles, but only two $6$-cycles after edge spreading.
To eliminate these $6$-cycles, assuming only one edge allocation can be changed at a time, the adjustable edge candidate set is $\{e_{v_4}^{c_1},e_{v_4}^{c_2},e_{v_4}^{c_3}\}$\footnote{For simplicity, we only consider the edges with the VN-to-CN direction.} and the corresponding initial spreading value set is $\{1,0,0\}$.
We modify each spreading value in turn, thus resulting in polynomials~$\{{\widetilde{p}_{\widetilde{\mathbf{T}}_{c_1,v_4}}^{6}},{\widetilde{p}_{\widetilde{\mathbf{T}}_{c_2,v_4}}^{6}},{\widetilde{p}_{\widetilde{\mathbf{T}}_{c_3,v_4}}^{6}}\}$ as in
\begin{equation}\label{eq:unigraph_modify_e}
	\left\{\!\!
	\begin{aligned}
		&\{0,0,0\}:\!\!\!\!&{\widetilde{p}_{\widetilde{\mathbf{T}}_{c_1,v_4}}^{6}}   &\!=\! \gamma_2^{\red{2}}\gamma_3^{\red{0}}\gamma_3^{\red{1}}+\gamma_1^{\red{-2}}\gamma_3^{-2}\gamma_3^{0}+\gamma_1^{\red{-1}}\gamma_1^{\red{0}}\gamma_2^0\gamma_2^2,\\
		&\{1,1,0\}:\!\!\!\!&{\widetilde{p}_{\widetilde{\mathbf{T}}_{c_2,v_4}}^{6}}   &\!=\! \gamma_2^{\red{2}}\gamma_3^{-1}\gamma_3^0\!+\!\gamma_1^{\red{-2}}\gamma_3^{\red{-3}}\gamma_3^{\red{-1}}\!+\!\gamma_1^{0}\gamma_1^1\gamma_2^{\red{1}}\gamma_2^{\red{3}},\\
		&\{1,0,1\}:\!\!\!\!&{\widetilde{p}_{\widetilde{\mathbf{T}}_{c_3,v_4}}^{6}}   &\!=\! \gamma_2^1\gamma_3^{\red{0}}\gamma_3^{\red{1}}\!+\!\gamma_1^{-1}\gamma_3^{\red{-1}}\gamma_3^{\red{1}}\!+\!\gamma_1^{\red{-1}}\gamma_1^{\red{0}}\gamma_2^{\red{-1}}\gamma_2^{\red{1}}.
	\end{aligned}
	\right.
\end{equation}
The difference between the original~$p_{v_4}^6$ and the updated polynomials is highlighted in red.
Hence, based on~\eqref{eq:unigraph_modify_e}, we can quickly compute the number of $6$-cycles containing~$v_{4}$ that can be removed as~$\{0,1,1\}$.
Note that despite one $6$-cycle removed by the modification of $\mathbf{T}_{2,4}$ from $0$ to $1$, it will inevitably lead to two new $4$-cycles, i.e., $f\left(\widetilde{p}_{\widetilde{\mathbf{T}}_{c_2,v_4}}^4\right)=f\left(\gamma_2^{0}+\gamma_1^{0}\gamma_3^{0}+\gamma_2^{0}\right)=(1+2+1)/2=2$.
Therefore, for~$v_4$, the modification of~$\mathbf{T}_{3,4}$ from~$0$ to~$1$ is the best choice to eliminate the maximum number of $6$-cycles without creating new~$4$-cycles.

Based on the aforementioned method, we use Remark~\ref{remark:unified_globaloptimal} to design a good spreading strategy on~$\mathcal{G}_{\UG}$.
\begin{remark}\label{remark:unified_globaloptimal}
	Given a target of eliminating $l$-cycles, when only one edge allocation is altered,~\eqref{eq:unigraph_argmax} provides the globally optimal to minimize the number of $l$-cycles in~$\mathcal{G}_{\UG}$, without introducing any new shorter cycles.
	\begin{equation}\label{eq:unigraph_argmax}
		\begin{aligned}
			\widetilde{\mathbf{T}}_{c,v}^{\ast}\triangleq\operatorname*{argmax}\limits_{\widetilde{\mathbf{T}}_{c,v}}&\underbrace{\left(f\left(p_v^{l}\right)-f\left(\widetilde{p}_{\widetilde{\mathbf{T}}_{c,v}}^{l}\right)\right)/2}_{\varepsilon_v^c}\\
			&\text{s.t.} \quad \widetilde{\mathbf{T}}_{c,v}\neq\mathbf{T}_{c,v}, \sum\limits_{j=4}^{l-2}Q_v^{j}=0.
		\end{aligned}
	\end{equation}
\end{remark}
\begin{proof}
	The metric~$\varepsilon_v^c$ in~\eqref{eq:unigraph_argmax} essentially measures the number of $l$-cycles containing a specific VN~$v$ that can be removed if reallocating the edge~$e_v^c$.
	Namely, we need to demonstrate that~${\varepsilon_v^c}$ is also the number of $l$-cycles removed in the entire~$\mathcal{G}_{\UG}$.
	First, assume that there exists a scenario where the total number of cycles eliminated in the whole graph is less than~${\varepsilon_v^c}$ (the proof follows similarly if it is `greater').
	For example, this edge allocation creates a new $l$-cycle that does not involve~$v$.
	It can be easily shown that this new cycle must include the edge~${e_{v}^{c}}$ since it is caused by this edge reallocation.
	Considering that an edge cannot exist independently of nodes, the edge~${e_{v}^{c}}$ must include~$v$, meaning this new cycle also involves~$v$.
	This contradicts our initial assumption.
	Hence, Remark~\ref{remark:unified_globaloptimal} always holds.
\end{proof}

\begin{algorithm}[t]
	\caption{Optimization on Unified Graph~$\mathcal{G}_{\UG}$}
	\label{alg:uigraph}
	\For{\upshape $i=0$ to $I_{\text{MP}}-1$}
	{
		\For{\upshape $v\in\mathbb{V}$}
		{ $p_v^{l}\leftarrow\mytextsf{MsgPass}(\mathcal{G}_{\UG},l);$\tcp{\!\!\!follow~\eqref{eq:unigraph_v2c}-\eqref{eq:unigraph_g2}}
		}
		$\widetilde{\mathbf{T}}_{c,v}^{\ast}$ 
		$\leftarrow$~\eqref{eq:unigraph_argmax};\\
		Update~$\mathcal{G}_{\UG}$ by~$\widetilde{\mathbf{T}}_{c,v}^{\ast}$;\\
	}
	$\textbf{return}\;\mathbf{T}$ and $\mathcal{G}_{\UG}$\;
\end{algorithm}
In Algorithm~\ref{alg:uigraph}, we summarize the approach to optimize the spreading strategy on the unified graph~$\mathcal{G}_{\UG}$, where~$I_{\text{MP}}$ is the maximum number of attempts for running message-passing (as well as the maximum number of edges that can be adjusted).
The operations from~\eqref{eq:unigraph_v2c} to~\eqref{eq:unigraph_g2} are summarized in the function~$\mytextsf{MsgPass}(\cdot)$.
Let~${e_{v}^{c}}^{\ast}$ and~${\varepsilon_v^c}^{\ast}$ denote the edge and the number of reduced~$l$-cycles, respectively, corresponding to the optimal~$\widetilde{\mathbf{T}}_{c,v}^{\ast}$.
As shown in~\eqref{eq:unigraph_argmax}, we adjust the edge~${e_{v}^{c}}^{\ast}$ with the allocation of~$\widetilde{\mathbf{T}}_{c,v}^{\ast}$ that achieves the greatest reduction in ${\varepsilon_v^c}^{\ast}$ $l$-cycles without any increase of shorter cycles.
Algorithm~\ref{alg:uigraph} is therefore a greedy algorithm to eliminate the residual cycles in~$\mathcal{G}_{\UG}$ in which each step identifies and modifies the edge that brings the steepest decline in cycle~count.

\subsubsection{Computational Complexity}
Following the extrinsic message-passing principle, the computational complexity of our algorithm is primarily attributed to the update of accumulated path values $\phi$ according to~\eqref{eq:unigraph_g1} and~\eqref{eq:unigraph_g2}.
For~$l$-cycles, the total complexity of cycle counting over~$\mathcal{G}_{\UG}$ is~$O(l\cdot|\mathbb{E}|\cdot|\mathcal{M}|)$, where $|\mathcal{M}|$ represents the average number of~$\gamma$ variables in a monomial.
This value is dynamic and upper bound by $\lceil 2Q_{v}^l/d_v\rceil$.
To eliminate~$l$-cycles on~$\mathcal{G}_{\UG}$, the overall complexity of~Algorithm~\ref{alg:uigraph} is~$O(I_{\text{MP}}\cdot\omega\cdot|\mathbb{E}|\cdot|\mathcal{M}|)$ when~$\omega$ is large.
However, if~$\omega\leq l$, this complexity will reduce to~$O(I_{\text{MP}}\cdot\l\cdot|\mathbb{E}|\cdot|\mathcal{M}|)$.
Namely, for a small~$\omega$, the additional complexity (line~$4$ of Algorithm~\ref{alg:uigraph}) introduced by this optimization on~$\mathcal{G}_{\UG}$ becomes negligible compared to the cycle counting part.

\subsection{Modified Coupled Reciprocal Channel Approximation}\label{subsec:RCA}
RCA~\cite{chung2000construction} is a fast and accurate approach to estimating the asymptotic iterative decoding threshold, which determines the minimum signal-to-noise-ratio (SNR) required for reliable transmission and has been successfully applied to the performance analysis of protograph-based LDPC ensembles under the AWGN channel~\cite{chen2015protograph}.
In RCA, the SNR, denoted as~$s$, is propagated along each edge of a protograph.
Unlike~\cite{chung2000construction} and~\cite{chen2015protograph}, we need to introduce the coupling length $L$ into the RCA analysis for ESRL codes.
Specifically, rather than using the uncoupled protomatrix~$\mathbf{B}$, we perform RCA on the coupled protomatrix~$\mathbf{B}_{\SC}$ to incorporate the effects of edge spreading and termination, i.e., including matrices~$\mathbf{T}$ and~$\mathbf{Q}$, thus enhancing the accuracy of the threshold estimation.
\begin{figure}[t]
	\centering
	\includegraphics[width=1\linewidth]{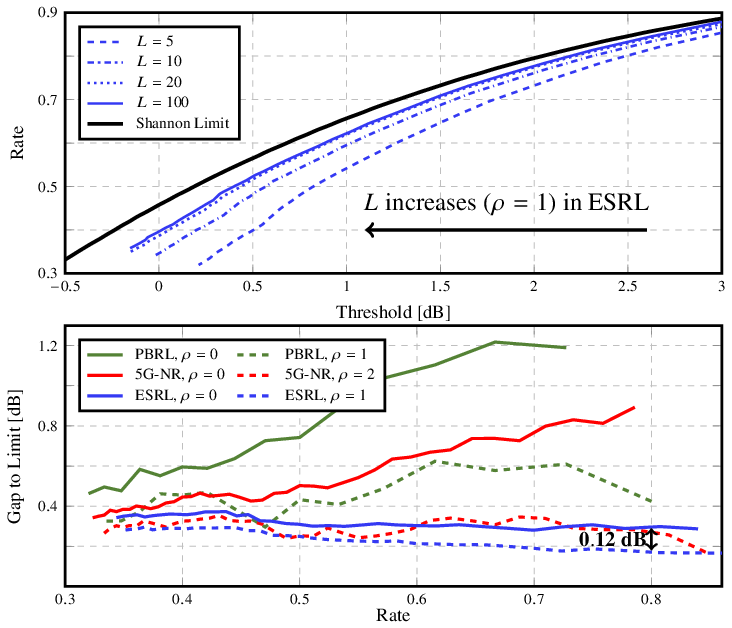}\\
	\caption{Threshold analysis comparing PBRL codes from~\cite{chen2015protograph}, 5G-NR LDPC codes from~\cite{5Gstandard2016}, and ESRL codes. In the lower figure, solid lines represent $\rho=0$ without puncturing, while dashed lines represent $\rho\neq 0$ with~puncturing.}\label{fig:RCA}
\end{figure}

Notably, the structured irregularity at the boundaries of SC-LDPC codes leads to a~\emph{wave-like} convergence, where the accumulated SNR of each bit progressively decreases from the ends toward the middle and the minimum accumulated SNR rises as $L$ increases~\cite{ranganathan2016some}. 
Considering the threshold is determined by the minimum SNR that satisfies $s>T$ for all VNs~\cite{chen2015protograph}, where $T$ is the stopping value in RCA, a limited~$L$ results in a higher estimated threshold.
Hence, the optimal threshold of an SC-LDPC code is achieved as~$L\!\rightarrow\!\infty$, aligning with the results of density evolution (DE)~\cite{nitzold2012spatially} (the gap to the limit vanishes as~$L$ increases).
However, as~$L$ grows, the~$\mathbf{B}_{\SC}$ size also increases significantly, making the RCA processing time-consuming~\cite{shi2022design}.
To address this issue, we present a modified coupled RCA tailored to ESRL codes to approach the threshold of~$L\!\rightarrow\!\infty$ with a finite~$L$ on~$\mathbf{B}_{\SC}$.

Let~$R(s)=C^{-1}(1-C(s))$ be the reciprocal energy function, where~$C(s)$ calculates the channel capacity under a given SNR (defined in~(5) of~~\cite{chen2015protograph}).
Denote the SNR message passed from the~$v$-th VN to the~$c$-th CN as~$s_{v,c}$, similarly, let~$s_{c,v}$ be the message passed from the~$c$-th CN to the~$v$-th VN.
At the~$t$-th iteration, the SNR messages are updated as
\begin{equation}\label{eq:RCA_update}
	\left\{
	\begin{aligned}
		s_{v,c}^{(t)}   &= s_{v,c}^{(0)} + \sum\limits_{c^{\prime}\in\mathbb{C}_v\backslash c} R\left(s_{c^{\prime},v}^{(t-1)}\right),\\
		s_{c,v}^{(t)}   &= \sum\limits_{v^{\prime}\in\mathbb{V}_c\backslash v} R\left(s_{v^{\prime},c}^{(t)}\right).
	\end{aligned}
	\right.
\end{equation}
The initialization~$s_{v,c}^{(0)}$ is set to input channel LLR if transmitted, or to~$0$ if punctured.
Hence, the overall reliability of each VN is computed as~$q_v^{(t)}=s_{v,c}^{(0)}+\sum_{c\in\mathbb{C}_v}R(s_{c,v}^{(t)})$, and~$\bm{q}^{(t)}$ thus forms a vector on~$\mathbf{B}_{\SC}$ across all~$(L\cdot n+\omega\cdot m)$ VNs of~$\mathbf{B}_{\SC}$.
In~\eqref{eq:RCA_averageL}, we first take the average of~$\bm{q}^{(t)}$ over the dimension of~$L$ spatial positions to mitigate the influence of a limited~$L$ on wave-like~convergence~\cite{ranganathan2016some}
\begin{equation}\label{eq:RCA_averageL}
	\bar{\bm{q}}^{(t)} = \frac{1}{L}\sum_{i=0}^{L-1} \bm{q}^{(t)}_{i\cdot n:(i+1)\cdot n-1},
\end{equation}
where $\bm{q}^{(t)}_{i\cdot n:(i+1)\cdot n-1}$ represents a segment of~$\bm{q}^{(t)}$ at spatial position~$i$, each of which has a length of~$n$.
Since the accumulated SNRs on the tail are lower relative to~$\bar{\bm{q}}^{(t)}$, to make~\eqref{eq:RCA_averageL} more accurate, we generate a weighted sum~$\widehat{\bm{q}}^{(t)}$ as
\begin{equation}\label{eq:RCA_weightedSum}
	\widehat{\bm{q}}^{(t)} = \frac{L\cdot n}{L\cdot n+\omega\cdot m}\cdot\bar{\bm{q}}^{(t)} + \frac{\omega\cdot m}{L\cdot n+\omega\cdot m}\cdot \zeta^{(t)}_{\tail}.
\end{equation}
We define the tail offset as~$\zeta_{\tail}^{(t)}=\frac{1}{\omega\cdot m}\sum_{v=L\cdot n}^{L\cdot n+\omega\cdot m-1} q^{(t)}_v$.
Finally, after $I_{\RCA}$ iterations, the minimum SNR that ensures that every element of $\widehat{\bm{q}}^{(t)}$ exceeds $T$ is recognized as the decoding~threshold.
In Fig.~\ref{fig:RCA}, the ESRL code designed in Section~\ref{sec:design_example} enjoys better thresholds across a wide range of rates compared to~5G-NR~\cite{5Gstandard2016}.
Besides, the puncturing scheme introduced in Section~\ref{subsec:puncture} can reduce thresholds by~$0.12$~dB at high rates for the ESRL code, which will be discussed later.
\begin{figure}[t]
	\centering
	\includegraphics[width=0.9\linewidth]{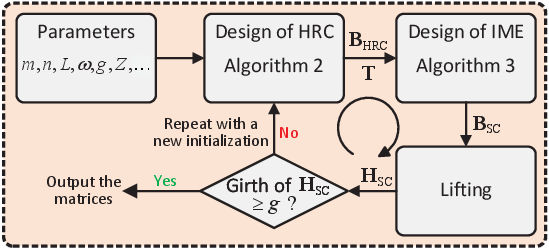}\\
	\caption{Design flow of the ESRL code.}\label{fig:designFlow}
\end{figure}

\subsection{Design Flow of ESRL Codes}\label{subsec:designFlow}
In this section, we present a comprehensive design flow for the proposed ESRL code.
In Fig.~\ref{fig:designFlow}, the process begins with basic parameters:~$\{m^{\prime},n^{\prime},m,n,\rho,\omega,L,g,Z,M,N\}$.
Herein, $m^{\prime}\times n^{\prime}$ represents the size of~$\mathbf{B}_{\HRC}$, while other parameters have been previously introduced, e.g., $M=\left(m-\rho\right)\cdot\left(L+\omega\right)\cdot Z$ and~$N=\left(n\cdot L+m\cdot\omega-\rho\cdot(L+\omega)\right)\cdot Z$ refer to~\eqref{eq:Rsc}.
The core of the design flow is divided into three procedures:~\circled[0.8]{1} Design of the HRC at a fixed high rate~\circled[0.8]{2} Design of the incremental matrix extension (IME), which integrates the IRC and SSC~\circled[0.8]{3} Circulant lifting to achieve a target code length and meet the desired girth~$g$.
\begin{algorithm}[t]
	\caption{Design of the HRC}
	\label{alg:designHRC}
	\begin{flushleft}
		\vspace{0.1em}
		\noindent \hspace*{-0.5em}\textbf{Step 1:} \texttt{Find the optimal~$p_{d_v}^*$ by DE} \\
		\vspace{0.1em}
		\noindent \hspace*{-0.5em}\textbf{Step 2:} \texttt{Generate $\mathbf{B}_{\HRC}$ through PEG + ACE} \\
		\noindent \hspace*{1em}$\mathbf{B}_{\HRC}\leftarrow\mytextsf{PEG}(p_{d_v}^\ast,m^\prime,n^\prime)$; \\
		\noindent \hspace*{1em}\textbf{if} $\mytextsf{ACE}(\mathbf{B}_{\HRC})<\eta_{\ACE}$ \textbf{then}\\ 
		\noindent \hspace*{3em}\text{Repeat }\textbf{Step 2};\\ 
		\vspace{0.1em}
		\noindent \hspace*{-0.5em}\textbf{Step 3:} \texttt{Optimize over unified graph~$\mathcal{G}_{\UG}$} \\
		\noindent \hspace*{1em}$\mathbf{T}\leftarrow\mathbf{0}$; \texttt{//initialization}\\
		\noindent \hspace*{1em}$\mathbf{T}\leftarrow\mathcal{G}_{\UG}(\mathbf{B}_{\text{HRC}},\mathbf{T})$; \texttt{//refer to Algorithm~\ref{alg:uigraph}}\\
		\vspace{0.1em}
		\noindent \hspace*{-0.5em}\textbf{Step 4}: \texttt{Perform modified coupled RCA} \\
		\noindent \hspace*{1em}$\mathbf{B}_{\text{SC}}\leftarrow\mytextsf{EdgeSpread}(\mathbf{B}_{\text{HRC}}, \mathbf{T}, \mathbf{Q}, L)$; \texttt{//as~\eqref{eq:Bsc}}\\ 
		\noindent \hspace*{1em}$\xi\leftarrow\mytextsf{RCA}(\mathbf{B}_{\text{SC}})$; \\
		\noindent \hspace*{1em}\textbf{if} $\xi<\xi^\ast$ \textbf{then}\\  
		\noindent \hspace*{3em}$\xi^\ast\leftarrow\xi,\;\mathbf{B}_{\text{HRC}}^\ast\leftarrow\mathbf{B}_{\text{HRC}},\;\mathbf{T^\ast\leftarrow\mathbf{T}}$;\\ 
		\vspace{0.1em}
		\noindent \hspace*{-0.5em}\text{Repeat} \textbf{Step 2} \text{to} \textbf{Step 4} $I_{\HRC}$ \text{times};\\
		\noindent \hspace*{-0.5em}\textbf{return} $\mathbf{B}_{\text{HRC}}^\ast$ and $\mathbf{T}^\ast$;
	\end{flushleft}
\end{algorithm}

The HRC design process is outlined in Algorithm~\ref{alg:designHRC}.
In~\textbf{Step~1}, under the ESRL code constraints from Section~\ref{sec:SecII-B}, DE is used to determine the optimal distribution of~$d_v$, denoted as the polynomial~$p_{d_{v}}^{\ast}$.
\textbf{Step 2} employs the progressive edge growth (PEG) algorithm~\cite{hu2005regular } and the approximate cycle extrinsic message degree (ACE) constraint~\cite{tian2004selective} to maximize local girth and to ensure the graph connectivity of short cycles on~$\mathbf{B}_{\HRC}$, respectively, which are two classical methods for designing finite-length LDPC codes.
Notably, threshold analysis on~$\mathbf{B}_{\HRC}$ is not part of our flow, aligning with the objective of ESRL codes in Section~\ref{sec:SecII-A} to focus on constructing a better coupled protomatrix~$\mathbf{B}_{\SC}$.
The next step is to use the unified graph~$\mathcal{G}_{\UG}$ in Section~\ref{subsec:unifiedGraph} and the modified coupled RCA in Section~\ref{subsec:RCA} to optimize~$\mathbf{B}_{\SC}$.
In~\textbf{Step 3}, the matrix~$\mathbf{T}$ is initially set to~$\mathbf{0}$, indicating that all edges are initially assigned to~$\mathbf{B}_{0}$.
The unified graph~$\mathcal{G}_{\UG}$ is then employed to optimize the spreading strategy, reallocating each edge to minimize cycle count.
In~\textbf{Step 4}, $\mathbf{B}_{\SC}$ is constructed using the code profile~$\mathbf{B}$,~$\mathbf{T}$, and~$\mathbf{Q}$, followed by modified coupled RCA to compute the threshold.
Since~$\mathbf{Q}$ has a fixed structure without additional design, it directly participates in the threshold analysis of~$\mathbf{B}_{\SC}$.
Finally, after~$I_{\HRC}$ iterations, the optimal~$\mathbf{B}_{\HRC}^{\ast}$ and~$\mathbf{T}^{\ast}$ with the lowest threshold are selected.

The IME design process is summarized in Algorithm~\ref{alg:designIME}.
Compared to the overall design of the HRC in Algorithm~\ref{alg:designHRC}, the IME is constructed row-wise until the full matrix of size~$m\times n$ is completed.
In the ESRL code, we find that the performance of the growing IME matrix is highly sensitive to the previously designed sections, especially after edge spreading, i.e., overemphasizing the performance of a specific intermediate rate will lead to significant performance degradation in subsequent matrix extensions.
To address this issue, instead of using a greedy approach to explore a local optimum for each newly added row as in~\cite{chen2015protograph}, we start with a function called~$\mytextsf{GlobalWeightRef}(\cdot)$ in \textbf{Step 1}, which essentially runs Algorithm~\ref{alg:designHRC} to design a reference~$\mathbf{B}$ of size~$m\times n$ directly and thus offers a preliminary IME example.
Note that although this reference is not optimal, it can provide a rough weight guideline~$d_{\IME}$ for the IME design to avoid being trapped in local optima for intermediate rows.
Similarly, \textbf{Step 2} involves using the PEG and ACE algorithms for the newly added row in~$\mathbf{T}$, which is initially set to zero.
In \textbf{Step 2.2}, the unified graph~$\mathcal{G}_{\UG}$ is generated to optimize the spreading strategy.
It is noteworthy that when designing the~$c$-th row of the IME, the edges and corresponding allocations of previous~$c-1$ rows must remain unchanged to preserve the optimization results obtained before.
Besides, certain special structures in ESRL codes, such as assigning the diagonal of the SSC to~$\mathbf{B}_{0}$, need to be retained.
Finally, we run the process~$I_{\IME}$-times for each row to search for optimal~$\mathbf{B}^{\ast}$ and~$\mathbf{T}^{\ast}$ with the lowest threshold, until completing the entire ESRL~protomatrix~design~$\mathbf{B}_{\SC}$.
\begin{algorithm}[t]
	\caption{Design of the IME}
	\label{alg:designIME}
	\begin{flushleft}
		\vspace{0.1em}
		\noindent \hspace*{-0.5em}\textbf{Step 1:} \texttt{Initialization}\\
		\noindent \hspace*{1em}$\mathbf{B}^{\ast}\leftarrow\mathbf{B}_{\HRC}^{\ast}$;\\
		\noindent \hspace*{1em}\texttt{//Avoid~trapping~into~local~optima}\\
		\noindent \hspace*{1em}$d_{\IME}\leftarrow\mytextsf{GlobalWeightRef}(\mathbf{B}_{\HRC},m,n)$;\\
		\vspace{0.1em}
		\noindent \hspace*{-0.5em}\textbf{Step 2:} \texttt{Generate IME row by row} \\
		\noindent \hspace*{1em}\textbf{for} $c=m^{\prime}$ \textbf{to} $m-1$ \textbf{then}\\
		\noindent \hspace*{-0.5em}\textbf{Step 2.1:} \texttt{Add a row through PEG + ACE} \\
		\noindent \hspace*{3em}$\mathbf{B}\leftarrow\mathsf{PEG}(\mathbf{B}^{\ast}, d_{\IME})$;\\
		\noindent \hspace*{3em}\textbf{if} $\mytextsf{ACE}(\mathbf{B})<\eta_{\ACE}$ \textbf{then}\\ 
		\noindent \hspace*{5em}\text{Repeat }\textbf{Step 2.1};\\ 
		\vspace{0.1em}
		\noindent \hspace*{-0.5em}\textbf{Step 2.2:} \texttt{Optimize over unified graph~$\mathcal{G}_{\UG}$}\\
		\noindent \hspace*{3em}$\mathbf{T}\leftarrow\mytextsf{AppendOneEmptyRow}(\mathbf{T}^{\ast})$; \texttt{//initial}\\
		\noindent \hspace*{3em}\texttt{//No changes in previous edges}\\
		\noindent \hspace*{3em}$\mathbf{T}\leftarrow\mathcal{G}_{\UG}(\mathbf{B},\mathbf{T})$;\\
		\vspace{0.1em}
		\noindent \hspace*{-0.5em}\textbf{Step 2.3:} \texttt{Perform modified coupled RCA}\\
		\noindent \hspace*{3em}$\mathbf{B}_{\text{SC}}\leftarrow\mytextsf{EdgeSpread}(\mathbf{B}, \mathbf{T}, \mathbf{Q}, L)$; \texttt{//as~\eqref{eq:Bsc}}\\ 
		\noindent \hspace*{3em}$\xi\leftarrow\mathsf{RCA}(\mathbf{B}_{\text{SC}})$; \\
		\noindent \hspace*{3em}\textbf{if} $\xi<\xi^*$ \textbf{then}\\  
		\noindent \hspace*{5em}$\xi^*\leftarrow\xi,\;\mathbf{B}^*\leftarrow\mathbf{B},\;\mathbf{T^*\leftarrow\mathbf{T}}$;\\ 
		\noindent \hspace*{3em}\text{Repeat} \textbf{Step 2.1} to \textbf{Step 2.3} $I_{\text{IME}}$ \text{times};\\
		\vspace{0.1em}
		\noindent \hspace*{-0.5em}\textbf{return} $\mathbf{B}^*$ and $\mathbf{T}^*$;
	\end{flushleft}
\end{algorithm}

The last step in the ESRL design flow is to incorporate the circulant lifting to transform~$\mathbf{B}_{\SC}$ into~$\mathbf{H}_{\SC}$, eliminating residual short cycles to meet the desired girth~$g$.
We define a matrix~$\mathbf{P}$ of size~$m\times n$ to store cyclic-shifting values of each non-zero entry in~$\mathbf{B}$, where~$\mathbf{P}_{c,v}\in[0,Z-1]$.\footnote{Note that since the ESRL code has no parallel edges, the matrix~$\mathbf{P}$ can be converted to~$\mathbf{B}$ by binarization, i.e.,~$\mathbf{P}$ inherently contains all the information of~$\mathbf{B}$. In practice,~$\mathbf{B}$ can be replaced by~$\mathbf{P}$ in the ESRL code profile, and we will no longer distinguish these two matrices in the remainder of this paper.}
According to the \emph{Fossorier condition}~\cite{fossorier2004quasicyclic}, an~$l$-cycle existing in~$\mathcal{G}_{\UG}$ persists in~$\mathbf{H}_{\SC}$ only if the lifting path~$\varphi$ of this cycle is also zero, as~shown~below\footnote{For brevity, we skip the description of lifting on the tail matrix~$\mathbf{Q}$ and the corresponding part in~$\mathbf{H}_{\SC}$, as the lifting process is similar.}
\begin{equation}\label{eq:Fossorier}
	\varphi=\Delta \mathbf{P}_0+\Delta \mathbf{P}_1+\Delta \mathbf{P}_2+\ldots+\Delta \mathbf{P}_{l/2-1},
\end{equation}
where~$\Delta\mathbf{P}_{i}=\mathbf{P}_{c_i,v_i}-\mathbf{P}_{c_i,v_{\text{mod}(i+1,l/2)}}$.
Notably,~\eqref{eq:unigraph_TBC_T} and~\eqref{eq:Fossorier} include an identical structure, which means that the circulant lifting step can also be incorporated into the optimization over~$\mathcal{G}_{\UG}$.
A design example will be provided in Section~\ref{sec:design_example}.

\section{Puncturing Scheme and Decoding Algorithm}\label{sec:punctured_node}
In this section, we introduce the puncturing scheme, describe the integration in the IR-HARQ process, and explain high-throughput SLME decoding tailored to ESRL~codes.
\begin{figure}[t]
	\centering
	\includegraphics[width=\linewidth]{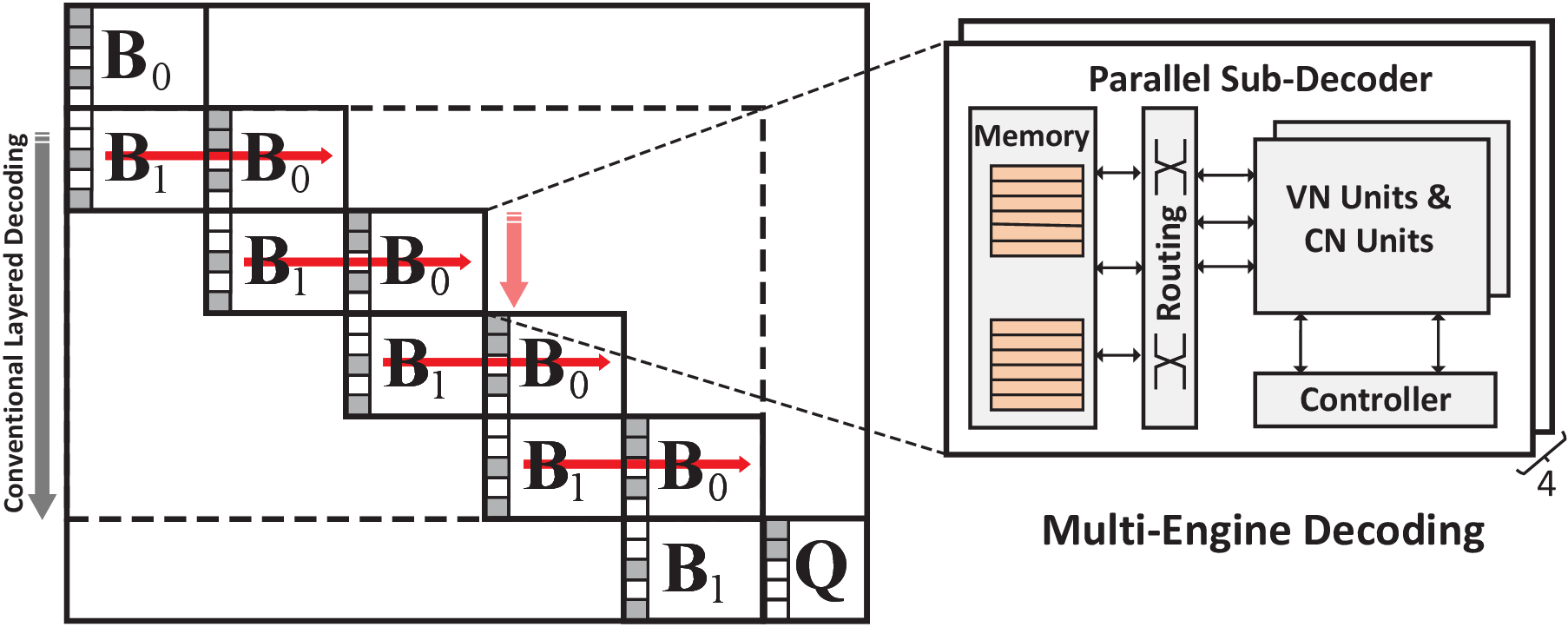}\\
	\caption{Puncturing scheme of the ESRL code, where the proposed semi-layered multi-engine decoding is running~\cite{ren2024asilomar}.}\label{fig:puncture}
\end{figure}

\subsection{Puncturing Scheme}\label{subsec:puncture}
Punctured bits can significantly reduce the decoding threshold and support higher transmission rates.
As shown in Fig.~\ref{fig:RCA}, both the PBRL~\cite{chen2015protograph} and 5G-NR LDPC codes~\cite{5Gstandard2016} benefit substantially from punctured bits.
In this section, we therefore also include puncturing in our ESRL codes.

As depicted in Fig.~\ref{fig:puncture}, the first VN at each spatial position of~$\mathbf{B}_{\SC}$ in the ESRL code is punctured, and these nodes are designed to connect with all CNs in the uncoupled protomatrix~$\mathbf{B}$ (described in Section~\ref{sec:SecII-B}).
Note that the first column of the tail matrix~$\mathbf{Q}$ is also punctured to maintain a regular structure.
This puncturing scheme improves the ESRL decoding threshold by~$0.12$~dB as visible in Fig.~\ref{fig:RCA}.
Moreover, an efficient puncturing scheme should also ensure good decoding convergence within a limited number of iterations~\cite{zhou2013robust}.
To evaluate this rapid recovery, we use the concept of a~\emph{one-step-recoverable VN}, as defined in~\cite{ha2004rate}.
Specifically, a punctured VN is called one-step-recoverable if it has at least one adjacent CN whose other VN neighbors are all unpunctured.
Notably,~the authors of~\cite{ha2004rate} proved that such a punctured VN can be fully recovered after the first decoding iteration on the BEC.
One-step-recoverable bits can also provide greater reliability on the AWGN channel compared to other types of punctured bits, leading to better performance with only a few iterations.
This leads to~Remark~\ref{remark:puncture}.
\begin{remark}\label{remark:puncture}
	All punctured VNs in the ESRL code belong to one-step-recoverable VNs, ensuring decoding convergence.
\end{remark}
\begin{proof}
	Given that the protomatrix~$\mathbf{B}$ is binary and contains no parallel edges, and only the first VN at each spatial position of~$\mathbf{B}_{\SC}$ is punctured, each CN thus connects to at most one punctured VN corresponding to an information bit.
	This is even stricter than the condition of one-step-recoverable VNs and Remark~\ref{remark:puncture}~always~holds.
	Therefore, our puncturing scheme not only optimizes the decoding threshold with higher transmission rates but also ensures good decoding convergence within a limited number of iterations.
\end{proof}

\subsection{IR-HARQ with ESRL Codes}\label{sec:support_ARQ}
\begin{figure}[t]
	\centering
	\includegraphics[width=\linewidth]{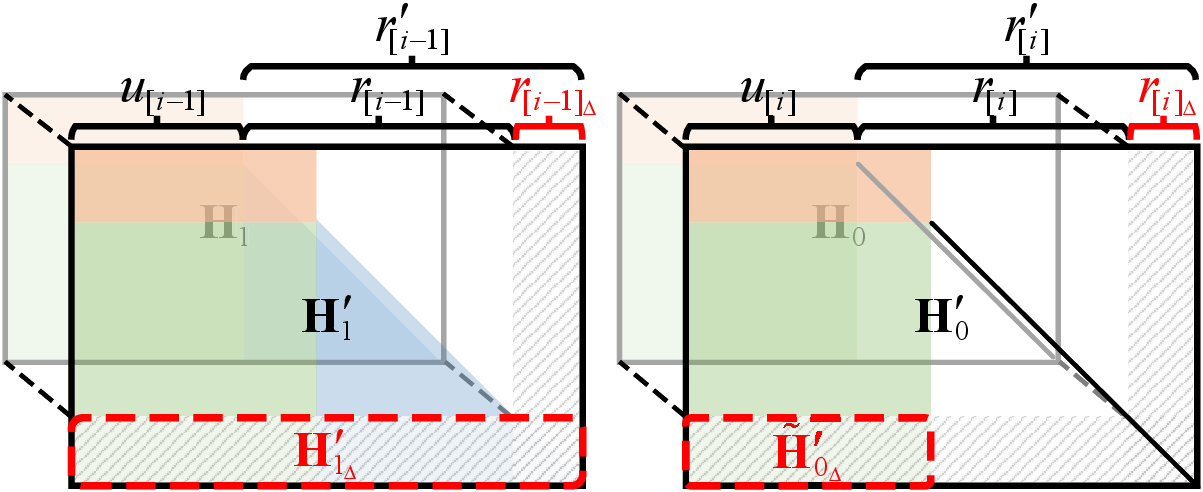}\\
	\caption{IR-HARQ with ESRL codes.}\label{fig:HARQ}
\end{figure}

We now demonstrate that ESRL codes can support the implementation of IR-HARQ.
When decoding an ESRL codeword~$\bm{c}_{[0,L-1]}$ fails, where $\bm{c}_{[i]}\triangleq[\bm{u}_{[i]},\bm{r}_{[i]}]$, sending~$L$ batches of additional parity redundancy~$\bm{r}_{{[i]}_{\Delta}}$ can still form a valid codeword of the ESRL code at a lower code rate.
The incremental codeword of each batch, $\bm{c}_{[i]}^{\prime}=[\bm{u}_{[i]},\bm{r}_{[i]}^{\prime}]=[\bm{u}_{[i]},\bm{r}_{[i]},\bm{r}_{{[i]}_{\Delta}}]$, combines to create~$\bm{c}_{[0,L-1]}^{\prime}=[\bm{c}_{[0]}^{\prime},\bm{c}_{[1]}^{\prime},\dots,\bm{c}_{[L-1]}^{\prime}]$, as shown~in~Fig.~\ref{fig:HARQ}.
\begin{proof}
	For brevity, we restrict~$\omega$ of ESRL codes to~$1$, while cases with~$\omega>1$ can be derived in a similar way.
	Based on SC-LDPC encoding~\cite{pusane2008implementation},~$\bm{r}_{[i]}=\widetilde{\mathbf{H}}_{0}\cdot\bm{u}_{[i]}+\mathbf{H}_{1}\cdot\bm{c}_{[i-1]},i\geq 1$, where~$\mathbf{H}_{0}$ is transformed as~$\mathbf{H}_{0}\triangleq[\widetilde{\mathbf{H}}_{0},\mathbf{I}]$.
	We assume the~$(i-1)$-th batch has already satisfied rate compatibility, i.e.,~$\bm{c}_{[i-1]}^{\prime}=[\bm{u}_{[i-1]},\bm{r}_{[i-1]}^{\prime}]=[\bm{u}_{[i-1]},\bm{r}_{[i-1]},\bm{r}_{{[i-1]}_{\Delta}}]$.
	Hence, in combination with Fig.~\ref{fig:HARQ}, we can obtain
	\begin{equation}\label{eq:H1C}
		\mathbf{H}_{1}^{\prime}\cdot\bm{c}_{[i-1]}^{\prime}=[\mathbf{H}_{1}\cdot\bm{c}_{[i-1]};\mathbf{H}_{{1}_{\Delta}}^{\prime}\cdot\bm{c}_{[i-1]}^{\prime}].
	\end{equation}
	Since~$\widetilde{\mathbf{H}}_{0}^{\prime}=[\widetilde{\mathbf{H}}_{0};\widetilde{\mathbf{H}}_{{0}_{\Delta}}^{\prime}]$ in Fig.~\ref{fig:HARQ}, we easily derive~$\widetilde{\mathbf{H}}_{0}^{\prime}\cdot\bm{u}_{[i]}=[\widetilde{\mathbf{H}}_{0}\cdot\bm{u}_{[i]};\widetilde{\mathbf{H}}_{{0}_{\Delta}}^{\prime}\cdot\bm{u}_{[i]}]$.
	Hence,~$\bm{r}_{[i]}^{\prime}$ of the~$i$-th batch after incremental redundancy transmission is~computed~as~\eqref{eq:HARQ_prove}
	\begin{equation}\label{eq:HARQ_prove}
		\begin{aligned}
			\bm{r}_{[i]}^{\prime}&=\widetilde{\mathbf{H}}_0^{\prime}\cdot \bm{u}_{[i]}+\mathbf{H}_1^{\prime}\cdot \bm{c}_{[i-1]}^{\prime}\\
			&=[\widetilde{\mathbf{H}}_0\cdot \bm{u}_{[i]}+\mathbf{H}_1\cdot c_{[i-1]};\widetilde{\mathbf{H}}_{0_\Delta}^{\prime}\cdot \bm{u}_{[i]}+\mathbf{H}_{1_\Delta}^{\prime}\cdot \bm{c}_{[i-1]}^{\prime}]\\
			&=[\bm{r}_{[i]};\widetilde{\mathbf{H}}_{0_\Delta}^{\prime}\cdot \bm{u}_{[i]}+\mathbf{H}_{1_\Delta}^{\prime}\cdot \bm{c}_{[i-1]}^{\prime}].
		\end{aligned}
	\end{equation}
	Herein, we can derive~$\bm{r}_{[i]}^{\prime}=[\bm{r}_{[i]},\bm{r}_{{[i]}_{\Delta}}]$ if~$\bm{r}_{{[i]}_{\Delta}}=\widetilde{\mathbf{H}}_{0_\Delta}^{\prime}\cdot \bm{u}_{[i]}+\mathbf{H}_{1_\Delta}^{\prime}\cdot \bm{c}_{[i-1]}^{\prime}$.
	Subsequently, this property can be recursively proven for all spatial positions~$i$,~$0\leq i< L$.
	As shown in~\eqref{eq:HARQ_final}, when decoding of~$\bm{c}_{[0,L-1]}$ fails, the incremental codeword~$\bm{c}_{[0,L-1]}^{\prime}$ remains valid if only sending additional parity redundancy bits.
	\begin{equation}\label{eq:HARQ_final}
		\setlength{\jot}{-2pt}
		\begin{aligned}
			&\bm{c}_{[0,L-1]}^{\prime}\!=\![\bm{u}_{[0]},\underbrace{\bm{r}_{[0]},\bm{r}_{[0]_\Delta}}_{\bm{r}_{[0]}^{\prime}},\ldots,\!\bm{u}_{[L-1]},\underbrace{\bm{r}_{[L-1]},\bm{r}_{{[L-1]}_\Delta}}_{\bm{r}_{[L-1]}^{\prime}}],\\
			&\quad\quad\quad\quad\quad\quad\quad\quad\quad\quad\quad\quad\uparrow \\
			&\quad\quad\quad\quad\bm{c}_{[0,L-1]}\!=\![\bm{u}_{[0]},\bm{r}_{[0]},\ldots,\bm{u}_{[L-1]},\bm{r}_{[L-1]}].\\
		\end{aligned}
	\end{equation}
\end{proof}
A comparison between ESRL codes and 5G-NR LDPC codes with respect to the performance under IR-HARQ is provided in~Section~\ref{sec:performance_HARQ}.

\subsection{SLME Decoding of ESRL codes}\label{subsec:decoding}
In this section, we propose a high-throughput semi-layered decoding algorithm using a multi-engine architecture in~\cite{schmalen2015spatially}, tailored to ESRL codes.
This algorithm can be flexibly configured as either high-performance full~BP decoding of the entire~$\mathbf{H}_{\SC}$ or as low-complexity windowed BP sliding along the diagonal of~$\mathbf{H}_{\SC}$.
Let~$W$ denote the window size,~$S$ the sliding step, and~$I_{\text{max}}$ the maximum number of decoding iterations within the window at each step.
For instance, by setting~$W\!=\!S\!=\!L+\omega$, SLME decoding corresponds to full~BP decoding with~$I_{\text{max}}$ for~ESRL~codes.

While coupling generally limits parallel decoding, ESRL codes without parallel edges can perform semi-layered decoding through a more compact multi-engine architecture compared to~\cite{schmalen2015spatially}, achieving higher parallelism.
Specifically, we apply individual layered decoding to each~$\mathbf{U}$ (defined in~\eqref{eq:ESRL_U} of Section~\ref{sec:SecII-A} and also shown in Fig.~\ref{fig:profile}), without causing memory read-write conflicts.
Two adjacent engines overlap and share the posterior LLR memory of~$\omega$ batches.
Thanks to the overlapping between engines, even with parallel processing across engines, each engine can still utilize the updated LLRs of the previous~$\omega$ batches provided by the preceding engines during layered decoding.
The `semi-layered' nature of SLME decoding therefore preserves the important data dependencies of ESRL codes, while the `multi-engine' approach significantly enhances decoding parallelism.
In Fig.~\ref{fig:puncture}, with a window size of~$W=4$, conventional layered decoding proceeds row-wise from top to bottom (indicated by a grey arrow).
In contrast, SLME decoding can deploy four identical sub-decoders working in parallel (each running layered decoding shown by a light red arrow in Fig.~\ref{fig:puncture}), thus resulting in a~$4\times$ throughput gain in~the~example.

It is noteworthy that since posterior LLRs are updated across multiple rows simultaneously in multi-engine, it highlights the importance of static decoding schedule optimizations for ESRL codes.
Inspired by~\cite{ren2024generalized}, we introduce a simple static schedule optimization that prioritizes rows with smaller~$d_c$, while postponing those with higher degrees to later in the sequence.
This priority ensures that reliable messages propagate to more VNs.
We define a layer ordering~$\mathbb{O}$ as
\begin{equation}\label{eq:SLME_order}
	\mathbb{O}\triangleq\{o_i,0\leq i<m|d_{c_{o_0}}\leq d_{c_{o_1}}\leq\ldots\leq d_{c_{o_{m-1}}}\}.
\end{equation}
The following simulation results for ESRL codes are based on this ordering~$\mathbb{O}$.
As mentioned before, setting~$W=S=L+\omega$ transforms SLME decoding into traditional BP decoding on~$\mathbf{H}_{\SC}$.
This configuration allows for a fair comparison with 5G-NR LDPC codes under equivalent computational complexity, which is discussed in~Section~\ref{sec:design_example}.

\section{Design Example and Performance Comparisons}\label{sec:design_example}
In this section, we provide a design example of ESRL codes and conduct comprehensive comparisons with 5G-NR.
\begin{figure}[t]
	\begin{minipage}{\linewidth}
		\centering
		\includegraphics[width=0.85\linewidth]{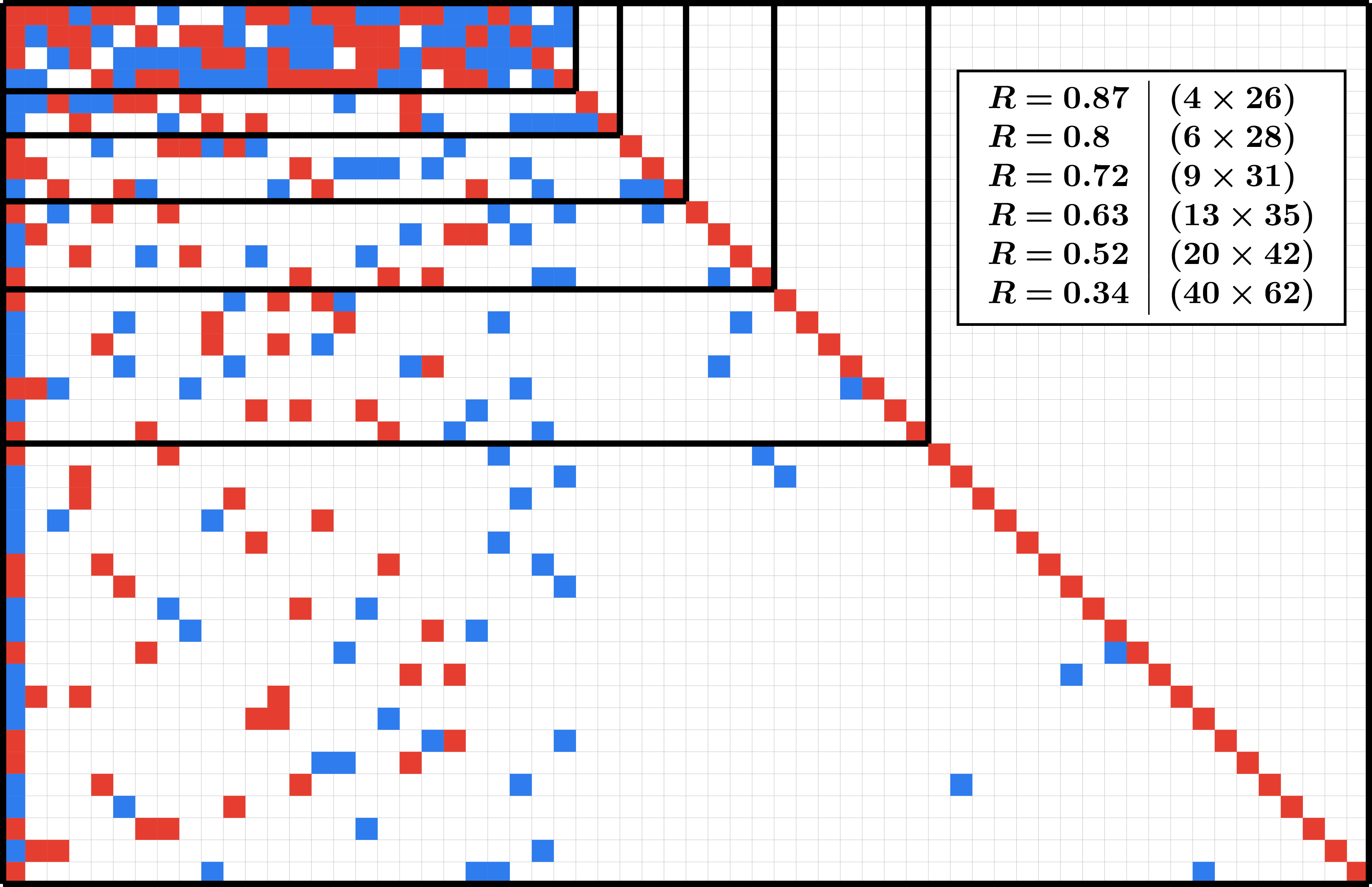}
	\end{minipage}\\[0.75em]
	\begin{minipage}{\linewidth}
		\centering
		\includegraphics[width=0.85\linewidth]{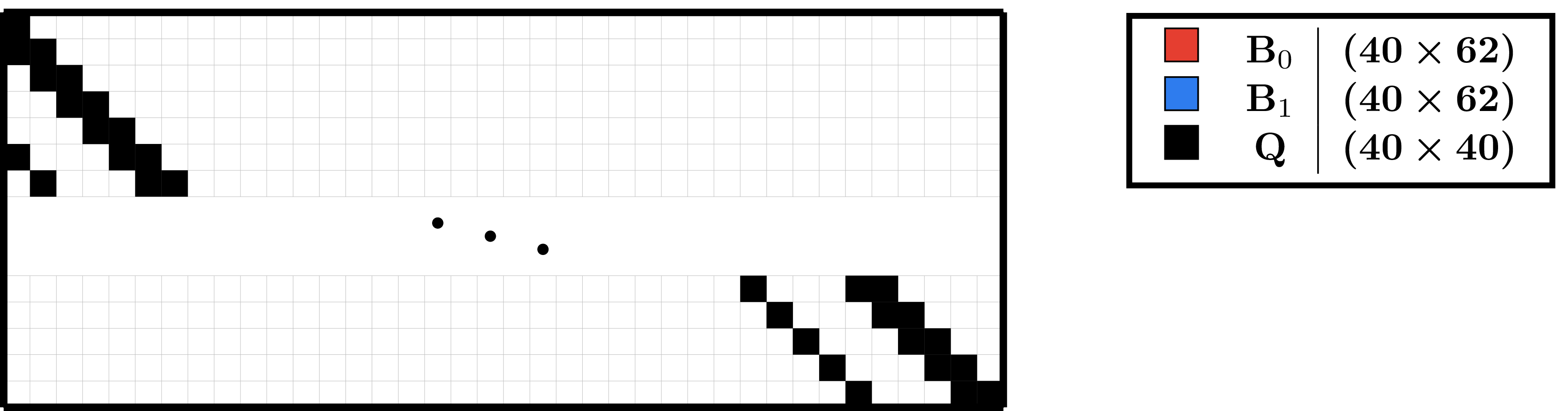}
	\end{minipage}
	\caption{Protomatrix example of ESRL codes is designed with~$k=22$ and~$\omega=1$, supporting a range of rates from~$0.87$ to~$0.34$ when~$L=10$.}\label{fig:exampleDesign}
\end{figure}
\begin{figure}[t]
	\centering
	\includegraphics[width=1\linewidth]{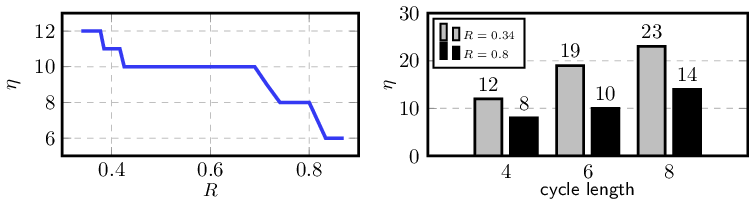}
	\caption{ACE spectrum of the protograph example of ESRL codes.}
	\label{fig:ACE}
\end{figure}

\subsection{Design Example of ESRL Codes}\label{sec:SecV_example}
\begin{figure}[t]
	\centering
	\includegraphics[width=\linewidth]{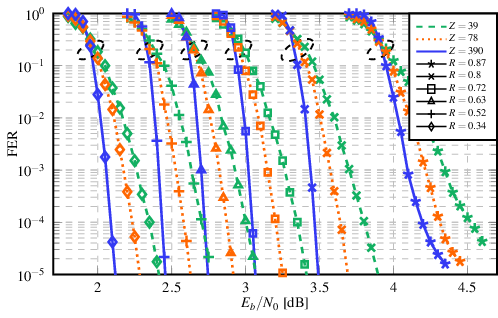}
	\caption{FER performance of ESRL codes with different lifting sizes.}
	\label{fig:FER_I5_ESRL_Z}
\end{figure}

\begin{figure*}[t]
	\centering
	\includegraphics[width=\linewidth]{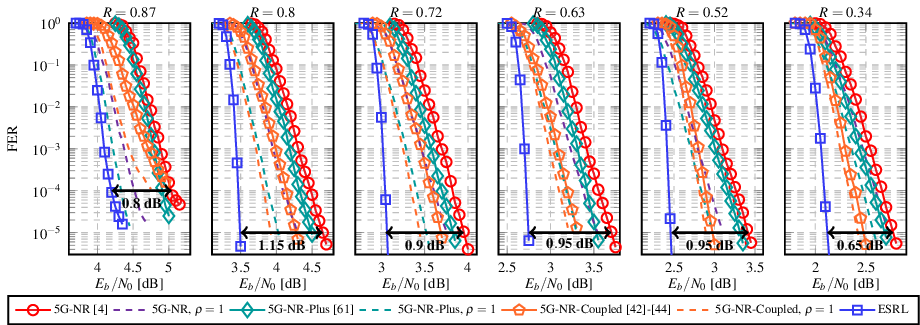}
	\caption{FER comparisons of ESRL codes and 5G-NR LDPC codes~\cite{5Gstandard2016} (including their variants~\cite{li2024efficient,nitzold2012spatially,wei2013coded,shi2022design}) using L-NMS decoding with~$I_{\text{max}}=5$.}
	\label{fig:FER_NMS_I5}
\end{figure*}

Fig.~\ref{fig:exampleDesign} illustrates an example ESRL protomatrix family designed for~$k=22$ (in each batch).
We intentionally set~$k$ equal to that of 5G-NR to facilitate fair comparisons in Section~\ref{sec:SecV_comp}.
Following SC-LDPC decoder implementations~\cite{SC_LDPC_database,griebel2023energy} to balance performance and complexity, we set~$\omega=1$ to simplify the code design.
Moreover, the coupling length~$L$ is set as~$10$ to balance the rate loss and the underlying blocklength for a given frame length.
Let~$m^{\prime}=4$,~$n^{\prime}=26$,~$m=40$, and~$n=62$.
Hence, the HRC protomatrix has dimensions of~$4\times 26$.
The IME consists of the IRC protomatrix (sized~$36\times 26$) and the SSC protomatrix (sized~$36\times36$). 
Considering that the first VN in the HRC is always punctured ($\rho=1$), the example ESRL code can support~$37$ distinct rates, ranging from the highest rate of~$0.87$ to the lowest rate of~$0.34$.
The uncoupled protomatrix~$\mathbf{B}$ has a size of~$40\times62$ and a weight of~$314$.
Note that with this configuration we mark six classical rates in Fig.~\ref{fig:exampleDesign}: $0.87~(4\times26)$,~$0.8~(6\times28)$,~$0.72~(9\times31)$,~$0.63~(13\times35)$,~$0.52~(20\times42)$, and~$0.34~(40\times62)$, which are also supported by 5G-NR and are used for the subsequent comparisons.
For instance, the highest rate is computed as~$R=(22\times 10)/(26\times10+4-1\times(10+1))\approx0.87$ using~\eqref{eq:Rsc}.
It is noteworthy that since the HRC is relatively dense, for~$R=0.87$, there exist some \emph{elementary trapping sets} with the number of VNs as~$2$ or~$3$, which explains the onset of the error floors in Fig.~\ref{fig:FER_I5_ESRL_Z} and Fig.~\ref{fig:FER_NMS_I5}.
The edge-spreading matrix~$\mathbf{T}$ is also reported in Fig.~\ref{fig:exampleDesign}, with~$\mathbf{B}_{0}$ and~$\mathbf{B}_{1}$ distinguished by colors.
As mentioned in Section~\ref{sec:SecII_codeprofile}, the tail matrix~$\mathbf{Q}$ has a fixed tri-diagonal structure, marked as black in Fig.~\ref{fig:exampleDesign}.
In terms of connectivity of short cycles, in the lowest rate protograph, every cycle of length up to~$8$ is connected to at least~$12$ extrinsic CNs.
For~$4$-cycles, the value of ACE~$\eta$ ranges from~$6$ to~$12$.
Fig.~\ref{fig:ACE} also shows the ACE spectrum of~$6$-cycles and~$8$-cycles.
It is noteworthy that given these ACE values are obtained from the protograph before lifting, the actual graph connectivity will be even better in~$\mathbf{H}_{\SC}$ of the ESRL~code~after~lifting.

Fig.~\ref{fig:FER_I5_ESRL_Z} shows the FER performance of the example ESRL codes under different lifting sizes.
Let~$\mathcal{C}$ define a series of rate-compatible LDPC codes with~$K$ information bits, and~$\mathcal{C}(N,K)$ be one of its code patterns with a rate~$R=K/N$.
Based on~$Z\in\{39,78,390\}$, three series of ESRL codes are defined: $\mathcal{C}_{\SC}^{\text{I}}$~$(K=8580)$,~$\mathcal{C}_{\SC}^{\text{II}}$~$(K=17160)$, and~$\mathcal{C}_{\SC}^{\text{III}}$~$(K=85800)$, where $K\!=\!k\cdot L\cdot Z$.
The maximum frame lengths of~$\mathcal{C}_{\SC}^{\text{I}}$,~$\mathcal{C}_{\SC}^{\text{II}}$, and~$\mathcal{C}_{\SC}^{\text{III}}$ are~$N=25311$,~$N=50622$, and~$N=253110$ for the lowest rate of~$R=0.34$, respectively, which are approximately~$1\times$,~$2\times$, and~$10\times$ longer compared to 5G-NR LDPC codes with the same rate.\footnote{Note that in this paper, we only consider the maximum blocklength 5G-NR LDPC codes with $Z = 384$ to fully present the 5G-NR performance.} 
In SLME decoding, we let~$W=S=L+\omega=11$, thus~decoding is configured to run BP on the entire~$\mathbf{H}_{\SC}$ of ESRL codes.
Given that hardware implementations as in~\cite{ren2024generalized,Lee2022TCASI,mohr2024region} typically limit the iteration count to~$5$ or below to minimize decoding delay and enhance throughput, we set~$I_{\text{max}}=5$ to highlight the ESRL performance in these high-throughput scenarios, as shown~in Fig.~\ref{fig:FER_I5_ESRL_Z}.
At~$R=0.87$, since the ESRL code only uses the HRC part without involving the IME, a slight error floor appears at the FER of~$10^{-5}$.
At other rates, ESRL codes exhibit a FER performance without error floors and achieve stable performance gains when~$Z$ increases.
For instance,~$\mathcal{C}_{\SC}^{\text{III}}$ with~$Z=390$ outperforms~$\mathcal{C}_{\SC}^{\text{I}}$ with~$Z=39$ by~$0.35$~dB~when~$R=0.72$ at the FER of~$10^{-5}$.
\begin{figure}[t]
	\centering
	\includegraphics[width=\linewidth]{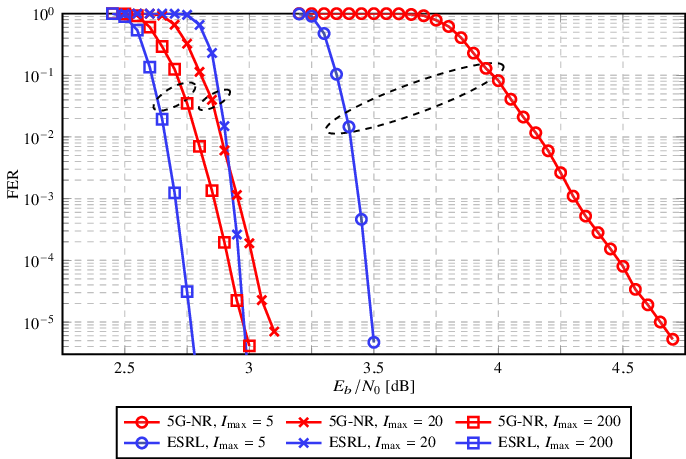}
	\caption{FER comparisons of ESRL codes and 5G-NR LDPC codes~\cite{5Gstandard2016} at~$R=0.8$ with different~$I_{\text{max}}$.}
	\label{fig:decodeI=5-20-200}
\end{figure}

\subsection{Comparisons with 5G-NR LDPC Codes}\label{sec:SecV_comp}
Comparing the performance of ESRL codes and 5G-NR LDPC codes in a reasonably fair manner involves many~factors.\footnote{Our decoding framework for 5G-NR LDPC codes is available at~\cite{nrLDPCRenGit}.} 
As stated in~\cite{zhang2023channel}, computational complexity does not always accurately reflect the true silicon area cost, and throughput is significantly affected by the hardware architecture and the level of processing parallelism.
Therefore, in addition to using the same code configurations, all decoding parameters and complexity should be the same in both software and~hardware as much as possible.

To concentrate on the selection of channel coding without the influence of decoding algorithms and hardware architectures, we adopt the following approach.
Specifically, we intentionally extend the uncoupled blocklength of ESRL codes to match that of 5G-NR LDPC codes. 
Combined with the multi-engine architecture in Section~\ref{subsec:decoding}, when SLME is configured as BP decoding on~$\mathbf{H}_{\SC}$, the corresponding ESRL decoder can be decomposed into~$L$ sub-decoders operating in parallel (with the tail redundancy causing some extra overhead).
Each sub-decoder is similar in size to that of a single 5G-NR LDPC decoder~\cite{ren2024generalized}.
In this context, compared to multi-core 5G-NR LDPC decoders~\cite{li2021high}, we can run the same decoding algorithm with the same hardware implementation for ESRL codes to eliminate additional~impacts.

For the comparisons, we select the ESRL code~$\mathcal{C}_{\SC}^{\text{III}}$, where each batch contains~$8580$ information bits, and the entire frame consists of a total of~$85800$ bits.
Meanwhile, we define~$\mathcal{C}_{\fiveG}^{\text{I}}$ as a packet frame that contains~$10$ individual 5G-NR LDPC codewords, with each code block containing~$8448$ information bits.
We employ a multi-core architecture to decode~$\mathcal{C}_{\fiveG}^{\text{I}}$, i.e., each sub-decoder independently works on an individual 5G-NR LDPC block.
It is noteworthy that each block in both~$\mathcal{C}_{\SC}^{\text{III}}$ and~$\mathcal{C}_{\fiveG}^{\text{I}}$ has a similar size, and their overall frame lengths are also comparable.
For each block in both the multi-engine architecture used for ESRL codes and the multi-core setup for 5G-NR LDPC codes, we employ the classical layered normalized min-sum (L-NMS) decoding and adopt the state-of-the-art block-parallel 5G-NR LDPC decoder in~\cite{ren2024generalized}.

For high-throughput scenarios with~$I_{\text{max}}=5$, as shown in Fig.~\ref{fig:FER_NMS_I5}, ESRL codes exhibit significant performance advantages compared to 5G-NR LDPC codes across various rates.
For instance, at~$R=0.8$, the ESRL code outperforms the 5G-NR LDPC code by~$1.15$~dB, and although the gain slightly decreases it remains~$0.65$~dB at~$R=0.34$.
Although 5G-NR LDPC is designed to puncture two columns, we change its~$\rho$ to~$1$ to enrich the comparison.
Under~$I_{\text{max}}=5$, the decoding convergence of~$\rho=1$ is better than that of~$\rho=2$ in 5G-NR LDPC codes, especially at~$R=0.87$.
Yet, 5G-NR LDPC codes with~$\rho=1$ are still worse than ESRL codes across various code rates.
Note that all algorithmic complexity and implementation complexity have been normalized, these gains are only attributable to channel coding.

Moreover, to fully present the FER performance of 5G-NR LDPC codes, we refer to~\cite{li2024efficient} and define such extended~5G-NR LDPC codes as~$\mathcal{C}_{\fiveG}^{\text{II}}$ $(K=84480)$ by redesigning a 5G-NR LDPC exponent matrix with a lifting size~$10\times$ larger $(Z=3840)$ and a non-decreasing girth.
The FER of $\mathcal{C}_{\fiveG}^{\text{II}}$, referred to as `5G-NR-Plus', is plotted in Fig.~\ref{fig:FER_NMS_I5}.
Given that the blocklength of the original 5G-NR LDPC codes is already on the order of tens of thousands, a~$10\times$ blocklength increase can bring only a~$0.15$ dB gain compared to the original 5G-NR LDPC codes.
ESRL codes still outperform 5G-NR-Plus LDPC codes~\cite{li2024efficient}.
Note that ESRL codes maintain better implementability than directly extending 5G-NR LDPC codes, since~$Z=3840$ will bring higher routing complexity to the shift network in decoders for an extended code~$\mathcal{C}_{\fiveG}^{\text{II}}$.
Subsequently, to intuitively highlight the difference between ESRL codes and previous protograph-based rate-compatible SC-LDPC code construction in~\cite{nitzold2012spatially,wei2013coded,shi2022design}, we replace the uncoupled~$\mathbf{B}$ of ESRL codes with the 5G-NR matrix and define the code~$\mathcal{C}_{\fiveG}^{\text{III}}$.
All other ESRL optimizations, such as the unified graph~$\mathcal{G}_{\UG}$ and the tail matrix~$\mathbf{Q}$, are still preserved in~$\mathcal{C}_{\fiveG}^{\text{III}}$ to ensure a fair comparison.
This more classical SC-LDPC construction refers to~`5G-NR-Coupled' in Fig.~\ref{fig:FER_NMS_I5}.
When taking~$R=0.63$ as an example, edge spreading for the 5G-NR matrix brings a~$0.4$~dB gain from coupling at the FER of~$10^{-5}$.
Even with one column punctured, compared~to the ESRL code, there is still a gap of~$0.45$~dB.
This demonstrates the significance of designing the optimal coupled matrix in ESRL codes, adhering to the design motivation behind~ESRL~codes.
\begin{figure}[t]
	\centering
	\includegraphics[width=\linewidth]{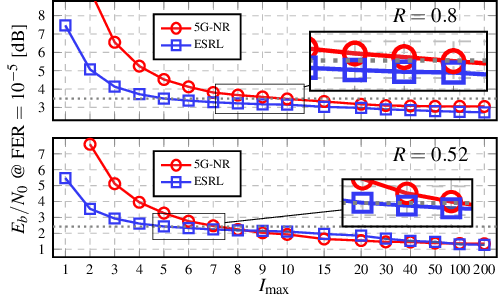}
	\caption{$E_b/N_0$ at the FER of $10^{-5}$ against~$I_{\text{max}}$.}
	\label{fig:SNRvsIter}
\end{figure}

\begin{figure}[t]
	\centering
	\includegraphics[width=\linewidth]{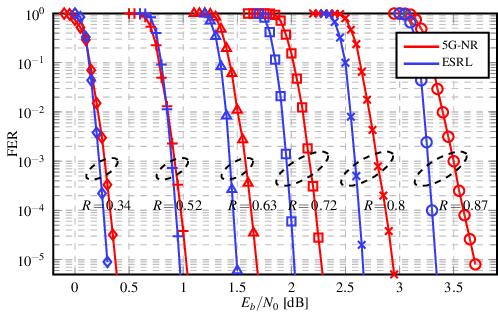}
	\caption{FER comparisons of ESRL codes and 5G-NR LDPC codes~\cite{5Gstandard2016} using L-SP decoding with~$I_{\text{max}}=200$.}
	\label{fig:FER_SP_I200}
\end{figure}

Fig.~\ref{fig:decodeI=5-20-200} highlights FER comparisons of ESRL codes and 5G-NR LDPC codes with different~$I_{\text{max}}$.
Due to the design of 5G-NR LDPC codes intended for~$20$ iterations, 5G-NR LDPC codes indeed perform better than ESRL codes in such cases.
However, in other scenarios of both a limited ($I_{\text{max}}=5$) and a large number of iterations ($I_{\text{max}}=200$),  ESRL codes show good performance over 5G-NR LDPC codes.
Fig.~\ref{fig:SNRvsIter} illustrates the relationship between $E_b/N_0$ at the FER of~$10^{-5}$ against~$I_{\text{max}}$ for ESRL codes and 5G-NR LDPC codes.
At~$R=0.52$, to match the ESRL performance of~$I_{\text{max}}=5$, the 5G-NR LDPC code needs to enhance $I_{\text{max}}$ to~$7$.
This gap widens further when~$R=0.8$.
We summarize all comparisons of ESRL codes with 5G-NR LDPC codes in Table~\ref{tab:hw_results}.
As shown in Table~\ref{tab:hw_results}, when~$I_{\text{max}}=5$, the ESRL decoder can deliver a peak throughput of~$209.4$~Gbps, while the 5G-NR LDPC decoder achieves a peak throughput of~$195.4$~Gbps.
These results are calculated by~(15) of~\cite{ren2024generalized}.
Notably, ESRL codes offer gains of more than~$0.8$~dB at most rates.
For 5G-NR with~$I_{\text{max}}=7$, although the FER performance becomes comparable, the peak throughput drops to~$139.5$~Gbps, indicating a~$50.1\%$ throughput improvement for ESRL codes.
For high-reliability scenarios, we use layered sum-product (L-SP) decoding~\cite{FR2001FGTIT} to replace L-NMS in Fig.~\ref{fig:FER_SP_I200}.
Under~$I_{\text{max}}=200$, ESRL codes outperform 5G-NR LDPC codes, especially at high code rates, e.g., this gain is~$0.36$~dB at the FER of~$10^{-5}$ when~$R=0.87$.
This greatly emphasizes the potential of ESRL codes for next-generation channel coding.
\begin{table}[t]
	\footnotesize
	\tabcolsep 1.8mm
	\renewcommand{\arraystretch}{1.2}
	\caption{Comparisons between ESRL codes and 5G-NR LDPC codes.}\label{tab:hw_results}
	\centering
	\begin{tabular}{V{2.5}lV{2.5}c|ccV{2.5}}
		\Xhline{1pt}
		\multicolumn{1}{V{2.5}cV{2.5}}{}       & \textbf{ESRL}           & \multicolumn{2}{cV{2.5}}{\textbf{5G-NR}}                    \\ \Xhline{1pt}
		\textbf{Max. Framelength}    & $253110$                          & \multicolumn{2}{cV{2.5}}{$253440$}                          \\ \hline
		\textbf{Max. Blocklength}    & $23790$                           & \multicolumn{2}{cV{2.5}}{$25344$}                           \\ \hline
		\textbf{Block Inf. Length}   & $8580$                            & \multicolumn{2}{cV{2.5}}{$8448$}                            \\ \hline
		\textbf{Architecture}        & Multi-Engine                      & \multicolumn{2}{cV{2.5}}{Multi-Core}                        \\ \hline
		\textbf{Decoder Cores}       & $10$                              & \multicolumn{2}{cV{2.5}}{$10$}                              \\ \hline
		\textbf{Algorithm}           & SLME (L-NMS)                      & \multicolumn{2}{cV{2.5}}{L-NMS}                             \\ \hline
		\textbf{Area}$^{\star}$      & $\approx 10\times$                & \multicolumn{2}{cV{2.5}}{$10\times$}                        \\ \hline
		\textbf{Routing}             & $Z=390$                           & \multicolumn{2}{cV{2.5}}{$Z=384$}                           \\ \Xhline{1pt}
		\textbf{Complexity}$^{\ast}$             & $I_{\text{max}}=5$    & \multicolumn{1}{c|}{$I_{\text{max}}=5$}      & $I_{\text{max}}=7$     \\ \hline
		\textbf{Peak T/P} [Gbps]$^{\dagger}$     & $209.4$               & \multicolumn{1}{c|}{$195.4$}                 & $139.5$                \\ \hline
		\textbf{FER}@$10^{-5}$$^{\ddagger}$      & $0.95$~dB~Better      & \multicolumn{1}{c|}{Baseline}                & $\approx$~ESRL         \\ \Xhline{1pt}
	\end{tabular}
	\begin{tablenotes}
		\scriptsize
		\item[*] $^{\star}$ Estimated by the posterior LLR memory overhead in hardware.
		\item[*] $^{\ast}$ Measured by the actual number of iterations running at each CN.
		\item[*] $^{\dagger}$ Peak throughput is attained with~$R=0.87$.
		\item[*] $^{\ddagger}$ Comparison is conducted at the FER of~$10^{-5}$ at~$R=0.52$.
	\end{tablenotes}
\end{table}

\subsection{Performance of Windowed BP}\label{sec:winBPfullBP}
\begin{figure}
	\centering
	\includegraphics[width=1\linewidth]{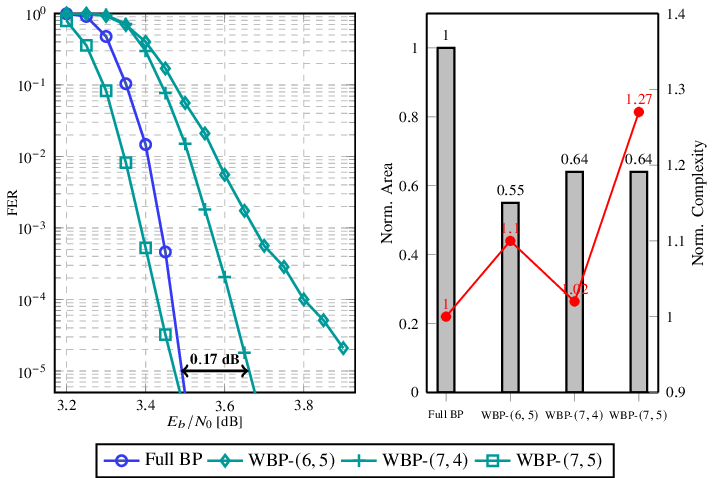}
	\caption{FER, normalized hardware area, and normalized computational complexity comparisons of windowed BP decoding and full BP decoding for ESRL codes with~$R=0.8$ and~$L=10$ under SLME decoding.}
	\label{fig:windowBP}
\end{figure}

ESRL codes can also support greater flexibility by windowed BP decoding~\cite{iyengar2011windowed}.
Fig.~\ref{fig:windowBP} presents a comparison between windowed BP with various configurations and full BP.
For brevity, let~WBP-$(W,I_{\text{max}})$ denote windowed BP decoding with a window size~$W$ and iteration count~$I_{\text{max}}$.
For instance, WBP-$(7,4)$ achieves a comparable computational complexity of full BP with~$I_{\text{max}}=5$. 
While this setting underperforms full BP by~$0.17$~dB at the FER of~$10^{-5}$, it reduces hardware area by approximately~$36.4\%$.
Moreover, by increasing the iteration count from~$I_{\text{max}}=4$ to~$I_{\text{max}}=5$, the performance of a decoder with this window size is even better than full BP with~$I_{\text{max}}=5$, due to the higher computational~complexity~of~WBP-$(7,5)$.
Note that since windowed BP utilizes less hardware and limits the message-passing to a smaller matrix dimension, its FER curve exhibits a slightly worse slope compared to full BP with the same number of iterations.
With the same computational complexity, full BP delivers the best performance but comes with the highest hardware consumption.
In contrast, windowed BP offers a broader design space exploration potential for ESRL codes, allowing optimization of parameters such as window size and sliding schedule to achieve a trade-off between performance and hardware complexity.

\subsection{Combination with IR-HARQ}\label{sec:performance_HARQ}
\begin{figure}[t]
	\centering
	\includegraphics[width=0.95\linewidth]{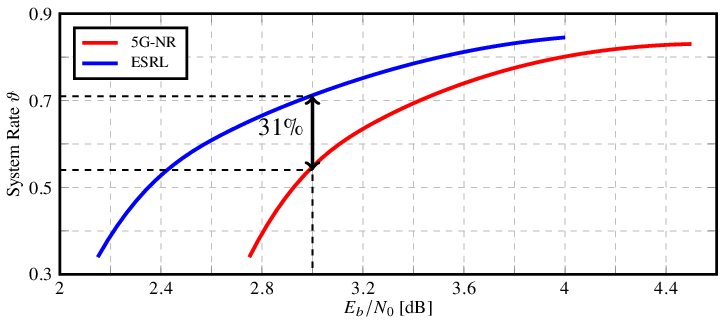}
	\caption{System rate comparison of ESRL codes and 5G-NR LDPC codes under IR-HARQ at a target FER of~$10^{-5}$ with~$I_{\text{max}}=5$.}
	\label{fig:HARQ_rate}
\end{figure}

Let~$\vartheta$ denote the system rate under IR-HARQ, i.e., defined as the average code rate of rate-compatible codes to satisfy a given frame error rate~(FER) performance (decoding starts from a high code rate), where more details are available in~(23) of~\cite{yue2007design}.
This metric can indicate the number of practically transmitted coded bits in wireless systems in an IR-HARQ process.
In combination with IR-HARQ, Fig.~\ref{fig:HARQ_rate} shows that at a target FER of~$10^{-5}$ using full BP decoding with~$I_{\text{max}}=5$, the system rate~$\vartheta$ of ESRL codes can be up to~$31\%$ higher than that of 5G-NR LDPC codes at $E_{b}/N_{0}=3.0$~dB.

\section{Conclusion}\label{sec:conclusion}
In this work, we propose an SC-LDPC code family called ESRL LDPC codes for 6G eMBB+ scenarios. 
The choice of an SC-LDPC code allows to extend and adjust the frame length by adjusting the coupling length in a flexible manner, without changing the matrix of the underlying block code. 
Larger block lengths thereby are not only in line with the desire to use larger transport block sizes in 6G eMBB+, but also enable high-throughput decoder architectures. 
To further enhance the throughput, the proposed ESRL code is optimized to not only achieve better error rate performance than 5G-NR LDPC codes with a high number of iterations, but also specifically with a low number of iterations.
This property is achieved by focusing on the coupled matrix in the code construction, rather than the base matrix only from which the SC-LDPC code is derived. 
As an essential feature to be of interest to 6G eMBB+, ESRL codes maintain rate compatibility with a Raptor-like structure that naturally evolves from the 5G-NR LDPC code and that maintains compatibility with the IR-HARQ process that is essential to~6G~eMBB+.

Further to the code construction and the proposal of a structure and code profile, we also present the complete design flow and optimization strategies, including a unified graph to optimize the edge spreading process and a modified coupled RCA to analyze the decoding threshold. 
To support the high-throughput requirement, we also propose a high-throughput decoding algorithm for ESLR codes, called SLME, which can be decomposed into multiple sub-decoders running in parallel to significantly enhance decoding parallelism. 
We provide a design example for a code and conduct comprehensive comparisons with 5G-NR LDPC codes and their variants, such as extended or coupled 5G-NR LDPC codes. 
When all decoding parameters and complexity are the same, the proposed code outperforms 5G-NR LDPC codes in terms of error rate and throughput in some specific scenarios (low and high number of iterations), which contributes a practical solution for applying SC-LDPC codes in future wireless systems.

\bibliographystyle{IEEEtran}
\bibliography{IEEEabrv, bibliography}

\end{document}